\documentclass[%
 reprint,
 amsmath,amssymb,amsthm,
 nofootinbib,
 aps,
]{revtex4-1}

\usepackage{dcolumn}
\usepackage{bm}
\usepackage[mathlines]{lineno}
\usepackage{graphicx}
\usepackage{csquotes}
\usepackage[final]{changes}

\begin{document}
\preprint{APS/123-QED}

\title{Alternative Framework to Quantize Fermionic Fields}

\author{Jianhao M. Yang}
\email[]{jianhao.yang@alumni.utoronto.ca}
\affiliation{Qualcomm, San Diego, CA 92121, USA}

\date{\today}		

\begin{abstract}
A variational framework is developed here to quantize fermionic fields based on the extended stationary action principle. From the first principle, we successfully derive the well-known Floreanini-Jackiw representation of the Schr\"{o}dinger equation for the wave functional of fermionic fields - an equation typically introduced as a postulate in standard canonical quantization. The derivation is accomplished through three key contributions. At the conceptual level, the classical stationary action principle is augmented to include a correction term based on the relative entropy arising from field fluctuations. Then, an extended canonical transformation for fermionic fields is formulated that leads to the quantum version of the Hamilton-Jacobi equation in a form consistent with the Floreanini-Jackiw representation; Third, necessary functional calculus with Grassmann-valued field variables is developed for the variation procedure. The quantized Hamiltonian can generate the Poincar\'{e} algebra, thus satisfying the symmetry requirements of special relativity. Concrete calculation of the probability of particle creation for the fermionic field under the influence of constant external field confirms that the results agree with those using standard canonical quantization. We also show that the framework can be applied to develop theories of interaction between fermionic fields and other external fields such as electromagnetic fields, non-Abelian gauge fields, or another fermionic field. These results further establish that the present variational framework is a novel alternative to derive quantum field theories. 
\end{abstract}

\maketitle
\section{Introduction}
In quantum field theory, there are two standard approaches to quantizing a classical field, both of which begin with formulating an appropriate Lagrangian density in terms of field variables. This Lagrangian may include interaction terms. The first approach, known as canonical quantization \cite{Weinberg, LancasterBook}, promotes the field variables and their conjugate momenta to operators and imposes commutation relations among them. The field operator is then expanded using creation and annihilation operators. The Fock space representation and the functional Schr\"{o}dinger representation are commonly used to describe the dynamics of field configurations. Perturbation theory is developed within the interaction representation. The second approach, path integral quantization \cite{FeynmanBook, Zee}, follows a more direct formulation. The action functional, obtained by integrating the Lagrangian density, is used to compute the probability amplitude for a given field configuration. By summing these amplitudes over all possible field configurations, one obtains the generating functional of the theory. Perturbation theory is then developed by expanding this functional in a series, allowing the derivation of various propagators. Both quantization approaches are complementary, offering different perspectives and techniques for quantum field theory.

In this paper, we propose an alternative mathematical framework for the second quantization of classical fields. This framework originates from the search for an information-theoretic foundation of quantum mechanics~\cite{Rovelli:1995fv, zeilinger1999foundational, Brukner:ys, Brukner:1999qf, Fuchs2002, brukner2009information, Paterek:2010fk, gornitz2003introduction, lyre1995quantum, Hardy:2001jk, Mueller:2012ai, Masanes:2012uq, chiribella2011informational, Mueller:2012pc, kochen2013reconstruction, 2008arXiv0805.2770G, Hall2013, Hoehn:2014uua, Hoehn:2015, Caticha2019, Frieden, Reginatto, Yang2021}, which led to the development of the extended principle of stationary action~\cite{Yang2023}. This extended principle has been shown to reproduce non-relativistic quantum mechanics for both spin-zero~\cite{Yang2023} and the spin-1/2 particle~\cite{Yang2025}. Given its broad applicability and the underlying mathematical structure, it is natural to extend this approach to field theory. Indeed, previous work has shown that scalar fields can be successfully quantized within this framework~\cite{Yang2024}. The goal of this paper is to formalize this alternative quantization framework and extend it to fermionic fields\footnote{Using information foundations to derive quantum field theory has recently become a research area of substantial interest. For example, an interesting theory of quantum gravity has been developed using an entropic action based on quantum relative entropy~\cite{Bianconi}. }.

A key step in the extended stationary action principle is the inclusion of an additional term in the Lagrangian that accounts for contributions from field fluctuations. This term is derived from the relative entropy, which quantifies the information distance between probability distributions with and without field fluctuations. By recursively applying the extended stationary action principle, we can determine the probability density of these fluctuations and derive the Schr\"{o}dinger equation for the wave functional of the fields. The general applicability of this principle stems from its foundation in the Lagrangian formalism, with the additional information-metric term incorporated via a general relative entropy formulation. However, new challenges arise when this approach is applied to the quantization of fermionic fields. Due to their inherently anticommutative nature, fermionic fields must be treated as Grassmann-valued variables, necessitating specialized mathematical techniques for defining inner products and performing integration by parts. In this paper, we derive a generalized Floreanini-Jackiw representation~\cite{Jackiw2, Kiefer} of the functional Schr\"{o}dinger equation for fermionic fields from first principles. This result is nontrivial, as it integrates multiple mathematical techniques, including the extended canonical transformation of classical fields, the use of Tsallis relative entropy for field fluctuations, and the variational calculus of functionals involving Grassmann variables.

Once the Schrödinger equation is derived and the Hamiltonian operator is obtained, we define the particle creation and annihilation operators, compute the vacuum state energy for fermionic fields, and more importantly, calculate the probability of particle creation when the fermionic field is influenced by a constant external electric field. These results are compared with the standard quantum field theory calculations and we find that they are in agreement. Additionally, we prove that the Hamiltonian operator, along with the momentum, angular momentum, and Lorentz boost operators, satisfies the Poincaré algebra. This confirms that the theory emerging from our quantization framework preserves the full symmetry structure required by special relativity.

Given the well-established success of canonical quantization and the path integral approach, both of which have been extensively verified experimentally, one may ask: what are the merits of this alternative quantization framework? This question can be addressed from both conceptual and mathematical perspectives. At the conceptual level, the extended stationary action principle provides a clear and intuitive explanation of how a classical field theory transitions into a quantum field theory. By incorporating information metrics that account for field fluctuations into the classical canonical framework, the theory naturally evolves into a fully quantum formulation. Introducing information metrics as a foundational element of quantum field theory represents a novel perspective that offers deeper insights into the role of information in quantum mechanics. At the mathematical level, the significance of this approach lies in its flexibility and broad applicability. It offers a potential alternative to conventional canonical quantization or the path integral formulation. For instance, we will demonstrate that this framework may be applied to quantize a non-renormalizable theory, leading to a nonlinear Schrödinger equation. Although our current formulation is developed in a Minkowski spacetime, extending it to a curved spacetime should be highly possible, which will be a topic for future research. 

The rest of the article is organized as follows. In Section \ref{sec:LIP}, we briefly review the underlying assumptions of the extended stationary action principle and formalize the step-by-step framework for quantizing a classical field. Sections \ref{sec:classicalTheory}, \ref{sec:extCanTrans}, and \ref{sec:QFT} apply this framework, recursively implementing the extended stationary action principle to derive the probability density of field fluctuations and ultimately obtain the generalized Floreanini-Jackiw representation of the Schrödinger equation for the wave functional of fermionic fields. The functional Hamiltonian operator is then verified to generate the Poincaré algebra in Section \ref{sec:Poincare}. To compare with the canonical quantization approach, in Section \ref{sec:comparison} we compute the probability of particle creation for the fermionic field under the influence of a constant external electric field, the result is consistent with the classical result of Schwinger~\cite{Schwinger}. In Section \ref{sec:interactions}, we extend the framework to include field interactions. Notably, we demonstrate that quantizing the non-renormalizable interaction between fermions leads to a nonlinear Schrödinger equation. Finally, Section \ref{sec:discussion} concludes the paper with a comparative analysis of different second quantization approaches and a discussion on the free parameters introduced in the present framework. Detailed mathematical techniques and derivations are provided in the Appendices.

\section{An Alternative Framework to Quantize a Classical Field}
\label{sec:LIP}
It has been shown that the principle of stationary action in classical mechanics can be extended to derive the theory of quantum scalar field by factoring in the following two assumptions~\cite{Yang2024}.
\begin{displayquote}
\emph{Assumption 1 -- There are constant fluctuations in the field configurations. The fluctuations are completely random and local.}
\end{displayquote}
\begin{displayquote}
\emph{Assumption 2 -- There is a lower limit to the amount of action that a physical system needs to exhibit in order to be observable. This basic discrete unit of action effort is given by $\hbar/2$ where $\hbar$ is the Planck constant.}
\end{displayquote}

The conceptual justifications of these two assumptions have been extensively discussed in Refs.\cite{Yang2023, Yang2024}. Assumption 2 provides us with a new way to calculate the additional action due to field fluctuations. That is, even though we do not know the physical details of field fluctuations, the field fluctuations manifest themselves via a discrete action unit determined by the Planck constant as an observable information unit. If we can define an information metric that quantifies the amount of observable information manifested by field fluctuations, we can then multiply the metric by the Planck constant to obtain the action associated with field fluctuations. Then, the challenge of calculating the additional action due to the field fluctuation is converted to define a proper new information metric $I_f$, which measures the additional distinguishable, and hence observable, information exhibited due to field fluctuations. The problem of defining an appropriate information metrics becomes less challenging since there are information-theoretic tools available. Information metrics that extract observable information about the dynamic effects of field fluctuations are defined by relative entropy. The concrete form of $I_f$ will be defined later as a functional of the Kullback-Leibler divergence $D_{KL}$, $I_f:=f(D_{KL})$, where $D_{KL}$ measures the information distances of different probability distributions caused by field fluctuations. Thus, the total action from classical path and vacuum fluctuation is
\begin{equation}
\label{totalAction}
    S_t = S_c + \frac{\hbar}{2}I_f,
\end{equation}
\added{where $S_c$ is the classical action. Quantum theory can be derived~\cite{Yang2024} through a variation approach to extremize such a functional quantity, $\delta S_t=0$. When $\hbar \to 0$, $S_t=S_c$. Extremizing $S_t$ is then equivalent to extremizing $S_c$, resulting in the classical dynamics for the fields. However, in quantum field theory, $\hbar \ne 0$, the contribution of $I_f$ must be included when extremizing the total action. We can see that the information metric $I_f$ is where the quantum behavior of a field comes from. These ideas can be condensed as\footnote{Along the development of this principle~\cite{Yang2023, Yang2024, Yang2025}, different names have been given to it, such as the principle of least observability, the extended principle of least action. The changes of name reflect the progressive understanding of the principle.}
\begin{displayquote}
\emph{\textbf{Extended Stationary Action Principle} -- The law of physical dynamics for a quantum field tends to extremize the action functional defined in (\ref{totalAction}).}
\end{displayquote}}

With this principle, the prescription for quantizing a classical field can be carried out with the following steps.

\begin{itemize}
    \item \textbf{Step I} Write down the Lagrangian density as that in the standard canonical quantization.
    \item \textbf{Step II} Apply the classical canonical transformation for the Lagrangian density such that the Hamilton-Jacobi equation is derived using the functional generator $S[\psi, t]$. To do this, we choose a foliation of the spacetime into a succession of spacetime hypersurfaces. Here we only consider the Minkowski spacetime, and it is natural to choose these to be the hypersurfaces $\Sigma_{t}$ of fixed $t$. Introducing the functional probability density $\rho[\psi, t]$ for an ensemble of field configurations in the hypersurface $\Sigma_{t}$, we can calculate the classical action $S_c$ for the ensemble of field configurations.
    \item \textbf{Step III} Apply the extended stationary action principle for an infinitesimal short time step. The relative entropy is defined as the information distance between the probability density due to field fluctuation and the complete uniform random probability density. Variation of the total action \eqref{totalAction} allows us to obtain the probability density of field fluctuation. From the probability density, we can calculate the variance of field fluctuations.
    \item \textbf{Step IV} Apply the extended stationary action principle again for a period of time to extract the equation for field dynamics. The key step here is to calculate the relative entropy as the information distance between $\rho[\psi, t]$ with and without field fluctuations in the hypersurface $\Sigma_{t}$. Summing the contributions to the relative entropy of all hypersurfaces $\Sigma_{t}$ for $t\in\{0, T\}$ results in an additional term $I_f$ in the Lagrangian density.
    \item \textbf{Step V} Carry out the variation procedure gives two differential equations for the dynamics of functional $S[\psi, t]$ and $\rho[\psi, t]$. Combining the two equations by defining the wave functional $\Psi=\sqrt{\rho}e^{iS/\hbar}$ gives the Schr\"{o}dinger equation for the wave functional.
    \item \textbf{Step VI} Verify that the Hamiltonian operator for the field dynamics can generate the Poincar\'{e} algebra. This step confirms that the theory satisfies the full symmetry required by special relativity\footnote{This step is not needed when the same framework is applied to derive the non-relativistic quantum mechanics, as shown in Ref.~\cite{Yang2023, Yang2025}.}.  
\end{itemize}
Steps I and II are still within the framework of classical field theory. The second quantization starts from Step III. 

Once the the Schr\"{o}dinger equation for the wave functional is derived and the correct Hamiltonian operator is identified, one can restore to the standard operator-based approach. For instance, operators for particle creation and annihilation can be defined, and the energy of the ground state or excited states can be calculated. The important point here is that the Schr\"{o}dinger equation is derived from the first principle rather than through a postulate in standard canonical quantization.

The framework ascribed above has been shown to successfully quantize the scalar fields~\cite{Yang2024}. For fermionic fields, additional challenges arise because the fields $\psi$ need to be considered as Grassmann variables and also have multiple components. We will develop the necessary mathematical tools to overcome these challenges and show that fermionic fields can be quantized using the same framework.


\section{The Lagrangian Density for Fermionic Fields}
\label{sec:classicalTheory}
Consider a massive fermionic field configuration $\psi$. Here we denote the coordinates for a four-dimensional spacetime point $x$ either by $x=(x^{(0)}, x^{(i)})$ where $i=\{1, 2, 3\}$. The field component at a spacetime point $x$ is denoted by $\psi_x=\psi(x)$. The standard Lagrangian density for the fermionic field is given by
\begin{equation}
\label{LD}
    \mathcal{L} = \bar{\psi}(i\gamma^\mu\partial_{\mu}-m)\psi.
\end{equation}
where $\mu=\{0, 1, 2, 3\}$ and the convention of Einstein summation is assumed. However, we will rewrite the Lagrangian density in an equivalent but more symmetric format
\begin{equation}
    \label{LD2}
    \begin{split}
        \mathcal{L} &= \frac{i}{2}\bar{\psi}\gamma^\mu\partial_\mu\psi - \frac{i}{2}\partial_\mu\bar{\psi}\gamma^\mu\psi - m\bar{\psi}\psi
    \end{split}
\end{equation}
Note that the field variables $\psi$ and $\psi^\dagger$ should be understood as variables with Grassmann values. From this Lagrangian density, the momentum conjugates to the fields $\psi$ and $\psi^\dagger$ are defined by
\begin{align}
    \label{momentum1}
    \pi_{\psi}(x) &= \frac{\delta\mathcal{L}}{\delta(\partial_0\psi)}=-\frac{i}{2}\psi^\dagger;\\
    \label{momentum2}
    \pi_{\psi^\dagger}(x) &= \frac{\delta\mathcal{L}}{\delta(\partial_0\psi^\dagger)}=-\frac{i}{2}\psi,
\end{align}
respectively. The minus sign in \eqref{momentum1} is due to the derivative of Grassmann variable. One can verify to obtain the Dirac equation from \eqref{LD2} through the Euler-Lagrange equation and
right multiplication of $\gamma^0$ on both sides of the Euler-Lagrange equation, resulting in
\begin{equation}
    \label{Dirac}
    (i\gamma^\mu\partial_\mu - m)\psi = 0.
\end{equation}
This confirms that the Lagrangians defined in \eqref{LD} and \eqref{LD2} are equivalent. However, \eqref{momentum1} and \eqref{momentum2} show that the field variables $\psi$ and $\psi^\dagger$ are on the equal footing if we use \eqref{LD2} as the Lagrangian. On the other hand, using \eqref{LD}, one will obtain $\pi_\psi=i\psi^\dagger$ and $\pi_{\psi^\dagger}=0$. We will choose \eqref{LD2} in subsequent formulations. 

Variables $(\psi, \pi_\psi)$ and $(\psi^\dagger, \pi_{\psi^\dagger})$ form two pairs of canonical variables, and the corresponding Hamiltonian is constructed by a Legendre transform of the Lagrangian
\begin{equation}
    \label{H1}
    \begin{split}
    &H[\psi, \pi_\psi, \psi^\dagger, \pi_{\psi^\dagger}] \\
    &=\int d^3x \{-\frac{i}{2}\psi^\dagger\gamma^0\gamma^i\partial_i\psi + \frac{i}{2}(\partial_i\psi^\dagger)\gamma^0\gamma^i\psi + m\psi^\dagger\gamma^0\psi\}
    \end{split}
\end{equation}
If we perform an integration by part for the second term, the Hamiltonian can be simplified as 
\begin{equation}
    \label{H2}
    \begin{split}
    H &= \int d^3x \{-i\psi^\dagger\gamma^0\gamma^i\partial_i\psi + m\psi^\dagger\gamma^0\psi\} \\
    &=\int d^3xd^3y\psi^\dagger(x) h(x,y)\psi(y),
    \end{split}
\end{equation}
where
\begin{equation}
    \label{h}
    h(x,y) = -i\gamma^0\gamma^i\partial_i\delta(x-y) + m\gamma^0\delta(x-y)
\end{equation}
is considered as the first quantized Dirac Hamiltonian. Eq.\eqref{H2} is the more familiar form of Hamiltonian appearing in the previous literature~\cite{Kiefer}. However, the Hamiltonian density in \eqref{H1}
\begin{equation}
    \label{HD}
    \mathcal{H} = -\frac{i}{2}\psi^\dagger\gamma^0\gamma^i\partial_i\psi + \frac{i}{2}(\partial_i\psi^\dagger)\gamma^0\gamma^i\psi + m\psi^\dagger\gamma^0\psi\
\end{equation}
has the advantage of treating $\psi$ and $\psi^\dagger$ on the equal footing. This property becomes important in the later development of our formulations.

\section{Extended Canonical Transformation}
\label{sec:extCanTrans}
Next step is to apply the canonical transformation technique in field theory. To do this, we will need to choose a foliation of the spacetime into a succession of spacetime hypersurfaces. Here we only consider the Minkowski spacetime and it is natural to choose these to be the hypersurfaces $\Sigma_{t}$ of fixed $t$. The field configuration $\psi$ for $\Sigma_{t}$ can be understood as a vector with infinitely many four-component spinors for each spatial point on the Cauchy hypersurface $\Sigma_t$ at time instance $t$ and denoted as $\psi_{t,\mathbf{x}}=\psi(t,\mathbf{x})$. For simplicity of notation, we will still denote $\psi(t,\mathbf{x})= \psi(x)$ for the rest of this paper, but the meaning of $\psi(x)$ should be understood as the field component $\psi_{\mathbf{x}}$ at each spatial point of the hypersurfaces $\Sigma_{t}$ at time instance $t$. We want to transform the pairs of canonical variables $(\psi, \pi_\psi)$ and $(\psi^\dagger, \pi_{\psi^\dagger})$ into generalized canonical variables $(\Phi, \Pi_\psi)$ and $(\Phi^\dagger, \Pi_{\psi^\dagger})$ and preserve the form of canonical equations. In Appendix \ref{appendix:canonical}, we show that by an extended canonical transformation, we have the following identifies
\begin{align}
\label{Conjugate}
    \frac{\delta S}{\delta\psi} &= \lambda\pi_\psi, \\
    \frac{\delta S}{\delta\psi^\dagger} &= \lambda\pi_{\psi^\dagger},
\end{align}
where $S(\psi, \psi^\dagger, t)$ is a generating functional, and $\lambda$ is a constant introduced in the canonical transformation. Substitute \eqref{momentum1} and \eqref{momentum2} into the above identities,
\begin{align}
\label{momemtum3}
    \frac{\delta S}{\delta\psi} &= -\frac{i}{2}\lambda\psi^\dagger, \\
\label{momemtum4}
    \frac{\delta S}{\delta\psi^\dagger} &= -\frac{i}{2}\lambda\psi.
\end{align}
The action functional after transformation is
\begin{equation}
\label{action2}
    A_c = -\int dt\{\frac{\partial S}{\partial t} + \lambda H[\psi, \pi, \psi^\dagger, \pi_{\psi^\dagger}]\}.
\end{equation}
Substituting \eqref{momemtum3}-\eqref{momemtum4} into $H$ in (\ref{H2}), we have
\begin{equation}
    \label{H3}
    H = -\frac{4}{\lambda^2}\int d^3xd^3y \{\frac{\delta S}{\delta\psi}h\frac{\delta S}{\delta\psi^\dagger} \}.
\end{equation}
A special solution to the stationary action principle based on the action functional in \eqref{action2} is $\partial S/\partial t + \lambda H = 0$, or,
\begin{equation}
    \label{HJE}
    \frac{\partial S}{\partial t} - \frac{4}{\lambda}\int d^3xd^3y \{\frac{\delta S}{\delta\psi}h\frac{\delta S}{\delta\psi^\dagger} \}= 0.
\end{equation}
This is the Hamilton-Jacobi equation for the fermionic field that governs the evolution of the functional $S$ among space-like hypersurfaces. It is equivalent to the Dirac equation in the Minkowski spacetime. 

However, there is a subtlety here. The variable $\psi$ can be interpreted as the field variable itself, or the momentum conjugate $\pi_{\psi^\dagger}$ due to \eqref{momentum2}. Therefore, for each $\psi$ in the Hamiltonian, there is a freedom to choose to leave it as is or to substitute it with $\delta S/\delta \psi^\dagger$ based on \eqref{momemtum4}. Similarly, $\psi^\dagger$ can be interpreted as the field variable itself or the momentum conjugate $\pi_{\psi}$ due to \eqref{momentum1}. For this reason, we find the Hamiltonian in \eqref{H1} to be more flexible. We can breakdown \eqref{H1} further into
\begin{equation}
\label{H4}
    \begin{split}
        H &=\int d^3x d^3y\{-\frac{i}{4}\psi^\dagger\gamma^0\gamma^i\partial_i\psi + \frac{i}{4}(\partial_i\psi^\dagger)\gamma^0\gamma^i\psi \\
        &-\frac{i}{4}\psi^\dagger\gamma^0\gamma^i\partial_i\psi + \frac{i}{4}(\partial_i\psi^\dagger)\gamma^0\gamma^i\psi + m\psi^\dagger\gamma^0\psi\}.
    \end{split}
\end{equation}
Then, for each of the first four terms, we use all combinations of leaving $\psi, \psi^\dagger$ as is or substituting with $\delta S/\delta \psi^\dagger$ or $\delta S/\delta \psi$, respectively. After integration by part, the resulting Hamiltonian is rewritten as
\begin{equation}
    \label{H5}
    H = \frac{1}{4}\int d^3xd^3y (\psi^\dagger + \frac{2i}{\lambda}\frac{\delta S}{\delta\psi})h(\psi+\frac{2i}{\lambda}\frac{\delta S}{\delta\psi^\dagger})
\end{equation}
The Hamilton-Jacobi equation becomes
\begin{equation}
    \label{HJE2}
    \frac{\partial S}{\partial t} + \frac{\lambda}{4}\int d^3xd^3y (\psi^\dagger + \frac{2i}{\lambda}\frac{\delta S}{\delta\psi})h(\psi+\frac{2i}{\lambda}\frac{\delta S}{\delta\psi^\dagger})= 0.
\end{equation}
Both \eqref{H3} and \eqref{H5} are valid Hamiltonian representations. They are equivalent since the corresponding original Hamiltonians before transformation, \eqref{H1} and \eqref{H2}, are equivalent through an integration by part.

Now we consider an ensemble of field configurations $(\psi, \psi^\dagger)$ in a hypersurface $\Sigma_t$. We assume that the ensemble follows a probability distribution with probability density $\rho[\psi, \psi^\dagger,t]$. Then, given \eqref{action2}, the action functional for the ensemble is
\begin{equation}
    \label{ensembleAction}
    S_c = -\int dt\mathcal{D}\psi^\dagger\mathcal{D}\psi \{\rho (\frac{\partial S}{\partial t} + \lambda H)\},
\end{equation}
where the Hamiltonian can be chosen from either \eqref{H3} or \eqref{H5}. Note that $S_c$ and $S$ are different functionals, where $S_c$ is the classical action functional of the ensemble, while $S$ is a generating functional introduced in the extended canonical transformation that satisfied the identities \eqref{momemtum3} and \eqref{momemtum4}. 

The pair of functionals $(\rho, S)$ can be treated as generalized canonical variables~\cite{Hall:2001, Yang2023, Yang2024, Yang2025}. when we apply the stationary action principle to the action functional defined in (\ref{ensembleAction}). Variation of $S_c$ with respect to $\rho$ leads to the Hamilton-Jacobi equation. Variation of $S_c$ with respect to $S$ gives an equation equivalent to the continuity equation for the probability density $\rho$, as shown later in Section V.B. The Hamilton-Jacobi equation and the continuity equation together determine the dynamics of the fermionic field ensemble before the second quantization.

\section{Second Quantization of the Fermionic Fields}
\label{sec:QFT}
\subsection{Probability Density of Field Fluctuations}
\label{sec:shorttime}
The first step in applying the extended stationary action principle is to investigate the dynamics of random fluctuations of the fermionic field. We consider such random field fluctuations in an equal-time hypersurface for an infinitesimal-time internal $\Delta t$. At a given time interval $t\to t+\Delta t$ in the hypersurface $\Sigma_t$, the field configurations fluctuate randomly, $\psi\to \psi + \omega$, $\psi^\dagger\to \psi^\dagger + \omega^\dagger$, where $\omega=\Delta\psi$ and $\omega^\dagger=\Delta\psi^\dagger$ are the changes of the field configurations due to random fluctuations. Define the probability density that the field configurations will transition from $\psi$ to $\psi + \omega$ and $\psi^\dagger$ to $\psi^\dagger + \omega^\dagger$ as $p[\omega, \omega^\dagger|\psi,\psi^\dagger]$. The action functional over all possible field fluctuations is 
\begin{equation}
    S_c=\int \mathcal{D}\omega^\dagger\mathcal{D}\omega \{\int dtd^3x p[\omega, \omega^\dagger|\psi,\psi^\dagger]\mathcal{L}[\psi+\omega, \psi^\dagger+\omega^\dagger]\}
\end{equation}
where $\mathcal{L}$ is given by (\ref{LD2}). Expanding (\ref{LD2}) explicitly with the derivatives of time and spatial variables, and noting that for an infinitesimal time internal $\Delta t$, one can approximate $\dot\psi=\omega/\Delta t$, and $\dot\psi^\dagger=\omega^\dagger/\Delta t$, we have
\begin{equation}
\begin{split}
    \mathcal{L}[\psi+\omega, \psi^\dagger+\omega^\dagger]&=\frac{i}{2}((\psi^\dagger+\omega^\dagger)\frac{\omega}{\Delta t}-\frac{\omega^\dagger}{\Delta t}(\psi+\omega)) \\&+ \frac{i}{2}(\psi^\dagger+\omega^\dagger)\gamma^0\gamma^i\partial_i(\psi+\omega) \\
    &- \frac{i}{2}\partial_i(\psi^\dagger+\omega^\dagger)\gamma^0\gamma^i(\psi+\omega) \\
    &-m(\psi^\dagger+\omega^\dagger)\gamma^0(\psi+\omega).
\end{split}
\end{equation}
By performing the integration over $\mathcal{D}\omega^\dagger\mathcal{D}\omega$ and noting the integration properties of the Grassmann variables, only those terms with both $\omega$ and $\omega^\dagger$ will survive, while any term with $\psi$ or $\psi^\dagger$ will vanish. We find
\begin{equation}
\label{action1}
\begin{split}
    S_c =& \Delta t\int d^3x\mathcal{D}\omega^\dagger\mathcal{D}\omega p[\omega, \omega^\dagger|\psi,\psi^\dagger]\\
    &\times(\frac{i}{2}\omega^\dagger\gamma^0\gamma^i\partial_i\omega - \frac{i}{2}\partial_i\omega^\dagger\gamma^0\gamma^i\omega-m\omega^\dagger\gamma^0\omega) \\
    =&-\Delta t\int d^3xd^3y\mathcal{D}\omega^\dagger\mathcal{D}\omega  p[\omega, \omega^\dagger|\psi,\psi^\dagger]\omega^\dagger h\omega.
\end{split}
\end{equation}
The last step uses the integration by part of the second term and $h$ is defined in \eqref{h}. Since the field variables $\psi$ and $\psi^\dagger$ vanishes in the integrand, we anticipate the transition probability density is independent of them, and can be simply denoted as $p[\omega, \omega^\dagger]$.

Next step is to define the information metric $I_f$ that is expected to capture the additional revelation of information due to field fluctuations in $\Sigma_t$. Thus, it is naturally defined as a relative entropy, or more specifically, the Kullback–Leibler divergence, to measure the information distance between $p[\omega, \omega^\dagger]$ and some prior probability distribution. Given that field fluctuations are completely random, it is intuitive to assume that the prior distribution is with maximal ignorance~\cite{Caticha2019, Jaynes}. That is, the prior probability distribution is a uniform distribution $\sigma$. 
\begin{align*}
    I_f  &:= D_{KL}(p[\omega, \omega^\dagger]|| \sigma) \\
    &= \int \mathcal{D}\omega^\dagger\mathcal{D}\omega p[\omega, \omega^\dagger]ln(\frac{p[\omega, \omega^\dagger]}{\sigma}).
\end{align*}
Combined with (\ref{action1}), the total action functional defined in (\ref{totalAction}) is (setting $\hbar=1)$
\begin{align*}
    S_t = &\int \mathcal{D}\omega^\dagger\mathcal{D}\omega\{ -p\Delta t\int d^3xd^3y( \omega^\dagger h\omega) + \frac{1}{2} pln\frac{p}{\sigma}\}.
\end{align*}
We also need to add the normalization constraint
\begin{equation}
    \label{constraint}
    \int \mathcal{D}\omega^\dagger\mathcal{D}\omega p[\omega, \omega^\dagger] =1.
\end{equation}
Taking the variation $\delta S_t = 0$ with respect to $p$ and with the above constraint through the Lagrangian multiplier method gives 
\begin{equation}
    \int\mathcal{D}\omega^\dagger\mathcal{D}\omega \delta p\{ -\Delta t\int d^3xd^3y \omega^\dagger h\omega+\frac{1}{2}ln\frac{p}{\sigma} +\frac{1}{2}+\lambda'\} = 0.
\end{equation}
where $\lambda'$ is the Lagrangian multiplier. Since $\delta p$ is arbitrary, one must have 
\begin{equation*}
    -\Delta t\int d^3xd^3y \omega^\dagger h\omega+\frac{1}{2}ln\frac{p}{\sigma} +\frac{1}{2} +\lambda'=0.
\end{equation*}
This gives the solution for $p$ as
\begin{equation}
\label{transP}
    p[\omega, \omega^\dagger] = \frac{1}{Z}\exp{(2\Delta t\int d^3xd^3y\omega^\dagger h\omega)},
\end{equation}
where $Z$ is a normalization factor $Z=\sigma^{-1} e^{1+2\lambda'}$. The value of $\lambda'$ can be determined by the normalization constraint. Equation (\ref{transP}) shows that the transition probability density for the field fluctuations in an infinitesimal time internal is a Gaussian-like distribution.

Recall that the fermionic field $\omega$ is a multi-component field. We label the components with indices $\alpha, \beta$, and adopt a compact notation
\begin{equation}
    \label{compact}
    \omega^\dagger h\omega \equiv \int d^3xd^3y\sum_{\alpha,\beta}\omega^\dagger_\alpha(x) h_{\alpha\beta}(x,y)\omega_\beta(y).
\end{equation}
Then \eqref{transP} can be written in a more compact form
\begin{equation}
\label{transP2}
    p[\omega, \omega^\dagger] = \frac{1}{Z}\exp\{2\Delta t(\omega^\dagger h\omega)\}.
\end{equation}
Given the probability density $p[\omega, \omega^\dagger]$, we want to calculate the expectation values $\langle\omega_\alpha\omega^\dagger_\beta\rangle$. However, the fermionic field components $\omega_\alpha$ and $\omega^\dagger_\beta$ are Grassmann variables. The inner product with Grassmann variables requires special treatment with the Berezin integral~\cite{Jackiw2, Kiefer}. In Appendix B, we show that with proper definition of inner product, the expectation value
\begin{equation}
\label{expectation}
    \langle\omega_\alpha(x)\omega^\dagger_\beta(y)\rangle = - h_{\alpha\beta}(x,y)\Delta t.
\end{equation}
On the other hand, due to the characteristics of Grassmann variables, it is straightforward to show that
\begin{equation}
\label{expectation2}
\begin{split}
    \langle\omega_\alpha(x)\rangle = \langle\omega^\dagger_\beta(y)\rangle &= 0,\\
    \langle\omega_\alpha(x)\omega_\alpha(y)\rangle &= 0,\\
    \langle\omega^\dagger_\beta(x)\omega_\beta^\dagger(y)\rangle &= 0.
\end{split}
\end{equation}
These properties are crucial in later calculations.

\subsection{The Functional Schr\"{o}dinger Equation for Fermionic Fields} 
\label{sec:SE}
We now turn to the field dynamics for a period of time from $t_A\to t_B$. As described earlier, the spacetime during the time duration $t_A\to t_B$ is sliced into a succession of $N$ Cauchy hypersurfaces $\Sigma_{t_i}$, where $t_i \in \{t_0=t_A, \ldots, t_i, \ldots, t_{N-1}=t_B\}$, and each time step is an infinitesimal period $\Delta t$. The field configuration for each $\Sigma_{t_i}$ is denoted as $\psi(t_i)$, which has an infinite number of components, labeled as $\psi_{\mathbf{x}}(t_i)=\psi(\mathbf{x}, t_i)$, for each spatial point in $\Sigma_{t_i}$. Without considering the random field fluctuation, the dynamics of the field configuration is governed by the Hamilton-Jacobi equation (\ref{HJE}), or \eqref{HJE2}. Furthermore, we consider an ensemble of field configurations for hypersurface $\Sigma_{t_i}$ that follow a probability density\footnote{The notation $\rho[\phi, t_i]$ is legitimate since in this case $\phi$ describes the field configuration for the equal time hypersurface $\Sigma_{t_i}$.} $\rho_{t_i}[\psi,\psi^\dagger] = \rho[\psi, \psi^\dagger, t_i]$. As mentioned in Section \ref{sec:classicalTheory}, the Hamilton-Jacobi equation and the continuity equation can be derived by the variation of the classical action functional $S_c$, as defined in (\ref{ensembleAction}), with respect to $\rho$ and $S$, respectively. 

To apply the extended stationary action principle, first we compute the action functional from the dynamics of the classical field ensemble as defined in (\ref{ensembleAction}). Next, we need to define the information metrics for the field fluctuations, $I_f$. For each new field configuration ($\psi+\omega$, $\psi^\dagger+\omega^\dagger$) where $(\omega, \omega^\dagger)$ are the change of field configuration due to field fluctuations between $t_i$ and $t_i+\Delta t$, there is a new probability density $\rho[\psi+\omega,\psi^\dagger+\omega^\dagger, t_i]$. Consequently, there is additional revelation of information due to the field fluctuations on top of the dynamics of the classical field ensemble. The proper measure of this distinction is the information distance between $\rho[\psi, \psi^\dagger, t_i]$ and $\rho[\psi+\omega,\psi^\dagger+\omega^\dagger, t_i]$. A natural choice for such an information measure is the relative entropy $D_{KL}(\rho[\psi, \psi^\dagger, t_i] || \rho[\psi+\omega,\psi^\dagger+\omega^\dagger, t_i])$. Moreover, we need to consider the contributions for all possible $\omega$. Thus, we take the expectation value of $D_{KL}$ over $\omega$ and $\omega^\dagger$, denoted as $\langle\cdot\rangle_{\omega}$. Then, the contribution of additional information due to field fluctuations in the hypersurface $\Sigma_{t_i}$ is $\langle D_{KL}(\rho[\psi, \psi^\dagger, t_i] || \rho[\psi+\omega,\psi^\dagger+\omega^\dagger, t_i])\rangle_{\omega}$. The expectation value should be evaluated with proper treatment of the Grassmann variable, as shown in Appendix \ref{appendix:exp}. Finally, we sum up the contributions from all the hypersurfaces, and obtain the definition of information metrics
\begin{align}
\label{DLDivergence}
    I_f &:= \sum_{i=0}^{N-1}\langle D_{KL}(\rho[\psi, \psi^\dagger, t_i] || \rho[\psi+\omega,\psi^\dagger+\omega^\dagger, t_i])\rangle_{\omega} \\
    &=\sum_{i=0}^{N-1}\langle\int \mathcal{D}\psi^\dagger\mathcal{D}\psi \rho [\psi, \psi^\dagger, t_i]ln \frac{\rho[\psi, \psi^\dagger, t_i]}{\rho [\psi+\omega, \psi^\dagger+\omega^\dagger,t_i]}\rangle_\omega.
\end{align}
With the detailed calculation shown in Appendix \ref{appendix:I_f}, we find that when $\Delta t\to 0$, $I_f$ turns out to be
\begin{equation}
\label{FisherInfo}
    I_f = \int dt\mathcal{D}\psi^\dagger\mathcal{D}\psi\int d^3xd^3y\frac{1}{\rho}\frac{\delta\rho}{\delta\psi(x)}h(x,y)\frac{\delta\rho}{\delta\psi^\dagger(y)}  .
\end{equation}
Eq. (\ref{FisherInfo}) is analogous to the Fisher information for the probability density in non-relativistic quantum mechanics~\cite{Yang2023, Frieden}. Some literature directly adds such Fisher information term in the variation method as a postulate to derive the Schr\"{o}dinger equation~\cite{Caticha2014, Ipek2021}. But (\ref{FisherInfo}) bears much more physical significance than the Fisher information. Defining $I_f$ using relative entropy opens up new results that cannot be obtained if $I_f$ is defined using Fisher information, because there are other generic forms of relative entropy such as R\'{e}nyi divergence or Tsallis divergence. 

Both \eqref{H3} and \eqref{H5} can be chosen as the Hamiltonian in the calculation of the action functional. We will start with \eqref{H3} first due to its simplicity, and study \eqref{H5} in Section \ref{sec:Jackiw}. Substituting \eqref{H3}, (\ref{ensembleAction}), and (\ref{FisherInfo}) into (\ref{totalAction}), we find that the total action functional is
\begin{equation}
    \label{totalDist}
    \begin{split} 
    S_t =& \int dt\mathcal{D}\psi^\dagger\mathcal{D}\psi\{-\rho\frac{\partial S}{\partial t} \\
    &+ \int d^3xd^3y (\frac{4\rho}{\lambda}\frac{\delta S}{\delta\psi}h\frac{\delta S}{\delta\psi^\dagger} + \frac{1}{2\rho}\frac{\delta\rho}{\delta\psi}h\frac{\delta\rho}{\delta\psi^\dagger})\}.
    \end{split}
\end{equation}
Performing variation of $S_t$ with respect to $S$ or $\rho$ is non-trivial due to the character of Grassmann variables. We need to develop the mathematical tool to carry out the integration by part for functional with Grassmann variables, as shown in Appendix \ref{Appendix:IntByPart}. Using these mathematical formulations, we show that variation of $S_t$ with respect to $S$ gives:
\begin{equation}
    \begin{split}
        \frac{\partial\rho}{\partial t} &- \frac{4}{\lambda}\int d^3xd^3y \{\frac{\delta \rho}{\delta\psi}h\frac{\delta S}{\delta\psi^\dagger} \\
        &+ \frac{\delta S}{\delta\psi}h\frac{\delta \rho}{\delta\psi^\dagger} + 2\rho\frac{\delta }{\delta\psi}h\frac{\delta S}{\delta\psi^\dagger}\}=0.
    \end{split}
\end{equation}
It can be written in a more compact form
\begin{equation}
\label{ContEq}
        \frac{\partial\rho}{\partial t} + \frac{4}{\lambda}\int d^3xd^3y \{\frac{\delta }{\delta\psi^\dagger}(\rho h^T\frac{\delta S}{\delta\psi}) -\frac{\delta }{\delta\psi}(\rho h\frac{\delta S}{\delta\psi^\dagger})  \}=0.
\end{equation}
This is the equivalence of the continuity equation for fermionic fields $\psi$ and $\psi^\dagger$. It can be derived by performing the variation procedure with the classical action functional defined in \eqref{action2}. Hence, \eqref{ContEq} is a classical result. There is no contribution of $I_f$ in the calculation of \eqref{ContEq} since $I_f$ is not dependent on $S$. On the other hand, variation of $I_f$ with respect to $\rho$ gives (see Appendix \ref{appendix:SE})
\begin{equation}
\label{varI_f}
    \delta'I_f = -\int dt\mathcal{D}\psi^\dagger\mathcal{D}\psi\int d^3xd^3y\{\frac{4}{R}\frac{\delta R}{\delta\psi}h\frac{\delta R}{\delta\psi^\dagger}\}\delta'\rho,
\end{equation}
where $R[\psi, \psi^\dagger,t]=\sqrt{\rho[\psi, \psi^\dagger, t]}$. Thus, variation of $S_t$ with respect to $\rho$ leads to
\begin{equation}
\label{QHJ}
\begin{split}
    \frac{\partial S}{\partial t} - \int d^3xd^3y\{\frac{\lambda}{4}\frac{\delta S}{\delta\psi}h\frac{\delta S}{\delta\psi^\dagger} - \frac{2}{R}\frac{\delta R}{\delta\psi}h\frac{\delta R}{\delta\psi^\dagger}\} =0,
    \end{split}
\end{equation}
This is the quantum version of the Hamilton-Jacobi equation for fermionic fields. The additional term in (\ref{QHJ}) compared to \eqref{HJE} is the fermionic field equivalence of the Bohm quantum potential~\cite{Bohm1952}. In non-relativistic quantum mechanics, the Bohm potential is considered responsible for the non-locality phenomenon in quantum mechanics~\cite{Bohm2}. Its origin is mysterious. Here we show that it originates from information metrics related to relative entropy, $I_f$. 

Define a complex functional 
\begin{equation}
\label{wavefunctional}
    \Psi[\psi, \psi^\dagger,t]=R[\psi, \psi^\dagger, t]\exp{(iS[\psi, \psi^\dagger, t])}.
\end{equation}
The continuity equation \eqref{ContEq} and the quantum Hamilton-Jacobi equation (\ref{QHJ}) can be combined into a single functional derivative equation when we choose $\lambda=2$ (see Appendix \ref{appendix:SE}),
\begin{equation}
    \label{SE}
    i\partial_0 \Psi = 2\{\int d^3xd^3y (\frac{\delta }{\delta\psi}h\frac{\delta}{\delta\psi^\dagger})\}\Psi.
\end{equation}
This is the Schr\"{o}dinger equation for the wave functional $\Psi[\psi, \psi^\dagger,t]$ with Hamiltonian operator density
\begin{equation}
    \label{operatorH}
    \hat{\mathcal{H}} = 2\frac{\delta }{\delta\psi}h\frac{\delta}{\delta\psi^\dagger}.
\end{equation}
It governs the evolution of the wave functional $\Psi[\psi,\psi^\dagger,t]$ between hypersurfaces $\Sigma_t$. 

The Schr\"{o}dinger equation in \eqref{SE} is different from the Floreanini-Jackiw representation of the Schr\"{o}dinger equation~\cite{Jackiw2}. This is due to the choice of the classical Hamiltonian \eqref{H3}. We will show in Section \ref{sec:Jackiw} that using the more general representation of the classical Hamiltonian \eqref{H5}, one can obtain the Floreanini-Jackiw representation of the Schr\"{o}dinger equation.

\subsection{Generalized Relative Entropy}

The term $I_f$ is supposed to capture additional observable information exhibited by field fluctuations and is defined in (\ref{DLDivergence}) as the summation of the expectation values of the Kullback-Leibler divergence between $\rho[\psi,\psi^\dagger,t]$ and $\rho[\psi+\omega,\psi^\dagger+\omega^\dagger, t]$. However, there are more general definitions of relative entropy, such as the Tsallis divergence~\cite{Renyi, Erven2014}. From an information-theoretic point of view, it is legitimate to consider alternative definitions of relative entropy. Suppose that we define $I_f$ based on Tsallis divergence\footnote{It is possible to choose other one-parameter generalization of relative entropy, such as R\`{e}nyi divergence. We expect that the result will be similar, as in the case of quantization of the scalar field \cite{Yang2024}. Tsallis divergence is chosen here because the subsequent calculations are simpler.},
\begin{align}
    &I_f^\alpha := \sum_{i=0}^{N-1}\langle D_T^\alpha(\rho[\psi, \psi^\dagger, t_i] || \rho[\psi+\omega,\psi^\dagger+\omega^\dagger, t_i])\rangle_{\omega} \\
    \label{TDivergence}
    &=\sum_{i=0}^{N-1}\langle\frac{1}{\alpha-1}(\int \mathcal{D}\psi^\dagger\mathcal{D}\psi \frac{\rho^\alpha[\psi, \psi^\dagger, t_i]}{\rho^{\alpha-1} [\psi+\omega, \psi^\dagger+\omega^\dagger,t_i]} - Z)\rangle_\omega.
\end{align}
The parameter $\alpha \in (0,1)\cup(1, \infty)$ is called the order of Tsallis divergence, and $Z = \int\mathcal{D}\psi^\dagger\mathcal{D}\psi\rho$ is an integration constant. Due to the characters of the Grassmann variables, it is not necessarily true that $Z=1$. The normalization factor $N$ for $\rho$ is defined in Appendix \ref{appendix:exp}. In Appendix \ref{appendix:RE}, we show that when $\Delta t\to0$,
\begin{equation}
\label{Tsallis}
    I_f = \alpha\int dt\mathcal{D}\psi^\dagger\mathcal{D}\psi\int d^3xd^3y\frac{1}{\rho}\frac{\delta\rho}{\delta\psi(x)}h(x,y)\frac{\delta\rho}{\delta\psi^\dagger(y)}.
\end{equation}
When $\alpha\to 1$, $I_f^{\alpha}$ converges to $I_f$ as defined in (\ref{FisherInfo}), as expected.

The parameter $\alpha$ provides a new degree of freedom to set the value of $\lambda$ when we derive the Schr\"{o}dinger equation. Using \eqref{Tsallis}, and following the same calculation steps in Appendix \ref{appendix:SE}, we find the quantum Hamilton-Jacobi equation becomes
\begin{equation}
\label{QHJ2}
\begin{split}
    \frac{\partial S}{\partial t} - \int d^3xd^3y\{\frac{\lambda}{4}\frac{\delta S}{\delta\psi}h\frac{\delta S}{\delta\psi^\dagger} -\frac{2\alpha}{R}\frac{\delta R}{\delta\psi}h\frac{\delta R}{\delta\psi^\dagger}\} =0.
    \end{split}
\end{equation}
By choosing $\alpha=2/\lambda$, the Schr\"{o}dinger equation takes the following form
\begin{equation}
    \label{SE2}
    i\partial_0 \Psi = \frac{4}{\lambda}\{\int d^3xd^3y (\frac{\delta }{\delta\psi}h\frac{\delta}{\delta\psi^\dagger})\}\Psi,
\end{equation}
and the Hamiltonian operator is 
\begin{equation}
    \label{H10}
    \hat{H}=\frac{4}{\lambda}\int d^3xd^3y (\frac{\delta }{\delta\psi}h\frac{\delta}{\delta\psi^\dagger}).
\end{equation}
Then \eqref{SE} is a special case of \eqref{SE2} when $\lambda=2$. Note that the choice of parameter $\lambda$ is constrained by the condition $\alpha > 0$.

\subsection{Floreanini-Jackiw Representation of Schr\"{o}dinger Equation}
\label{sec:Jackiw}

To derive the Floreanini-Jackiw representation of Schr\"{o}dinger Equation for fermionic field from the extended stationary action principle, we need to use the more general representation of Hamiltonian \eqref{H5}. Using \eqref{H5}, \eqref{ensembleAction}, and \eqref{Tsallis}, the total action functional is
\begin{equation}
    \label{totalAction4}
    \begin{split} 
    S_t =& \int dt\mathcal{D}\psi^\dagger\mathcal{D}\psi\{\int d^3xd^3y (\frac{\alpha}{2\rho}\frac{\delta\rho}{\delta\psi}h\frac{\delta\rho}{\delta\psi^\dagger})-\rho\frac{\partial S}{\partial t}\\
    &- \frac{\lambda\rho}{4}\int d^3xd^3y (\psi^\dagger + \frac{2i}{\lambda}\frac{\delta S}{\delta\psi})h(\psi+\frac{2i}{\lambda}\frac{\delta S}{\delta\psi^\dagger}) \}.
    \end{split}
\end{equation}
The mathematical procedure of variation using this action functional $S_t$ is more tedious but follows the same calculation steps as in Appendix \ref{appendix:SE}. Variation of $S_t$ over $S$ gives
\begin{align*}
    \label{contEqJackiw}
    &\frac{\partial\rho}{\partial t}+\int d^3x d^3y\{\frac{i}{2}(\frac{\delta\rho}{\delta\psi}h\psi + \psi^\dagger h\frac{\delta\rho}{\delta\psi^\dagger}) \\
    &-\frac{1}{\lambda}(\frac{\delta \rho}{\delta\psi}h\frac{\delta S}{\delta\psi^\dagger} + \frac{\delta S}{\delta\psi}h\frac{\delta \rho}{\delta\psi^\dagger} + 2\rho\frac{\delta }{\delta\psi}h\frac{\delta S}{\delta\psi^\dagger})\}=0.
\end{align*}
Variation of $S_t$ over $\rho$ gives the quantum Hamilton-Jacobi equation.
\begin{align*}
    \frac{\partial S}{\partial t} &+ \frac{\lambda}{4}\int d^3xd^3y (\psi^\dagger + \frac{2i}{\lambda}\frac{\delta S}{\delta\psi})h(\psi+\frac{2i}{\lambda}\frac{\delta S}{\delta\psi^\dagger})\\
    &+\int d^3x d^3y\frac{2\alpha}{R}\frac{\delta R}{\delta\psi}h\frac{\delta R}{\delta\psi^\dagger} = 0
\end{align*}
Defined the complex functional $\Psi[\psi,\psi^\dagger,t]$ as in \eqref{wavefunctional}, and set the parameter $\alpha\lambda=1/2$, the above two equations are combined into a single functional derivative equation
\begin{equation}
    \label{generalSE}
    i\partial_0\Psi = \{\frac{\lambda}{4}\int d^3xd^3y (\psi^\dagger + \frac{2}{\lambda}\frac{\delta }{\delta\psi})h(\psi+\frac{2}{\lambda}\frac{\delta }{\delta\psi^\dagger})\}\Psi,
\end{equation}
and the Hamiltonian operator is
\begin{equation}
    \label{generalHDO}
    \hat{H} = \frac{\lambda}{4}\int d^3xd^3y(\psi^\dagger + \frac{2}{\lambda}\frac{\delta }{\delta\psi})h(\psi+\frac{2}{\lambda}\frac{\delta }{\delta\psi^\dagger}), 
\end{equation}
where $\lambda > 0$ since the parameter $\alpha>0$. Equation (\ref{generalSE}) gives a family of linear functional derivative equations for each $\lambda$, and each $\lambda$ corresponds to each order of the Tsallis divergence. When $\lambda=2$, we obtain the well-known Floreanini-Jackiw representation of the functional Schr\"{o}dinger equation for fermionic fields,
\begin{equation}
    \label{JackiwSE}
    i\partial_0\Psi = \{\frac{1}{2}\int d^3xd^3y (\psi^\dagger + \frac{\delta }{\delta\psi})h(\psi+\frac{\delta }{\delta\psi^\dagger})\}\Psi,
\end{equation}
and
\begin{equation}
    \label{JackiwHDO}
    \hat{H} = \frac{1}{2}\int d^3xd^3y(\psi^\dagger + \frac{\delta }{\delta\psi})h(\psi+\frac{\delta }{\delta\psi^\dagger}).
\end{equation}

Once the Hamiltonian density operator is identified, standard operator-based quantum field theory can be applied, such as defining the particle creation and annihilation operators and calculating the typical results such as the vacuum energy, the probability of particle creation, as with the canonical quantization approach, as to be shown in next section.

Two comments are in order with respect to the derivation of the Floreanini-Jackiw representation of the functional Schr\"{o}dinger equation. First, since $\lambda=2$, we have $\alpha=1/4$. It is interesting that to derive the Floreanini-Jackiw representation of the functional Schr\"{o}dinger equation, we need to use the Tsallis divergence to define the information metrics $I_f$ and set $\alpha=1/4$. In fact, if we use the standard Kullback–Leibler divergence, we have $\alpha=1$ and thus $\lambda=1/2$, which results in the following form of functional Schr\"{o}dinger equation
\begin{equation}
    \label{DLSE}
    i\partial_0\Psi = \frac{1}{8}\{\int d^3xd^3y (\psi^\dagger + 4\frac{\delta }{\delta\psi})h(\psi+4\frac{\delta }{\delta\psi^\dagger})\}\Psi.
\end{equation}

Second, the Hamiltonian operator \eqref{generalHDO} is derived from the initial Hamiltonian \eqref{H1}. Although Hamiltonian \eqref{H1} is equivalent to Hamiltonian \eqref{H2}, after second quantization, we cannot interpret the Grassmann variable $\psi$ in \eqref{generalHDO} to be the same as that in \eqref{H2}. To avoid ambiguity, we denote the field variables in \eqref{generalHDO} with a different set of symbols $\{u(x), u^\dagger(y)\}$ instead of $\{\psi(x), \psi^\dagger(y)\}$, so that
\begin{equation}
    \label{generalHO2}
    \hat{H} = \frac{\lambda}{4}\int d^3xd^3y(u^\dagger + \frac{2}{\lambda}\frac{\delta }{\delta u})h(u+\frac{2}{\lambda}\frac{\delta }{\delta u^\dagger}), 
\end{equation}
One may argue that starting from the Hamiltonian \eqref{H2} and promoting the field variables in\eqref{H2} to operators as 
\begin{align}
\label{transform1}
    \psi &\to \frac{\sqrt{\lambda}}{2}(u+\frac{2}{\lambda}\frac{\delta}{\delta u^\dagger}),\\
\label{transform2}
    \psi^\dagger &\to \frac{\sqrt{\lambda}}{2}(u^\dagger+\frac{2}{\lambda}\frac{\delta}{\delta u}),
\end{align}
the Hamiltonian \eqref{H2} is quantized\footnote{Alternatively, by promoting $\psi\to \frac{2}{\sqrt\lambda}\frac{\delta}{\delta \psi^\dagger}$ and $\psi^\dagger\to \frac{2}{\sqrt\lambda}\frac{\delta}{\delta \psi}$, one quantizes the Hamiltonian \eqref{H2} to \eqref{H10} to \eqref{generalHO2}. This form of Hamiltonian operator is less preferred for reasons discussed later. In supersymmetric field theory~\cite{Duncan}, the transformation in \eqref{transform1} and \eqref{transform2} can be interpreted as relating the realization of field operators ($u, u^\dagger$) to the holomorphic representation ($\psi, \psi^\dagger$).}. In fact, this method is used in the standard canonical quantization~\cite{Jackiw2, Kiefer}. However, the above promotion appears rather ad hoc. Instead, the quantization method presented in the current paper clearly shows how the Hamiltonian operator \eqref{generalHDO} can be derived from the first principle, the extended stationary action principle.

In summary, by recursively applying the same extended stationary action principle in two steps, we recover the Schr\"{o}dinger representations of the standard relativistic quantum theory of fermionic fields~\cite{Jackiw2, Kiefer}. In the first step, we consider the dynamics of field fluctuations in a hypersurface $\Sigma_t$ for an infinitesimal short period of time interval $\Delta t$, and obtain the transitional probability density due to field fluctuations; In the second step, we apply the same principle for a cumulative time period to obtain the dynamics laws that govern the evolutions of $\rho$ and $S$ between the hypersurfaces. The applicability of the same principle in both steps shows the consistency and simplicity of the theory, although the forms of Lagrangian density are different in each step. In the first step, the Lagrangian density $\mathcal{L}$ is given by (\ref{LD2}), while in the second step, we use a different form of the Lagrangian density $\mathcal{L}^\prime = -\rho(\partial S/\partial t + H)$. As shown in Appendix \ref{appendix:canonical}, $\mathcal{L}$ and $\mathcal{L}^\prime$ are related through an extended canonical transformation. The choice of Lagrangian $\mathcal{L}$ or $\mathcal{L}^\prime$ does not affect the outcomes of the variation procedure, that is, the form of Legendre's equations. We choose $\mathcal{L}^\prime$ as the Lagrangian density in the second step in order to use the pair of functionals $(\rho, S)$ in the subsequent variation procedure.  


\section{Poincar\'{e} Group and Algebra}
\label{sec:Poincare}
It is important to note that the derivation of the Schr\"{o}dinger equation (\ref{JackiwSE}) depends on a particular foliation of Minkowski spacetime. Therefore, the theoretical framework presented here treats the time parameter differently and it is not obvious whether the theory is Lorentz invariant. To confirm that the theory is fully relativistic, one must verify that the Hamiltonian operator $\hat{H}$ given by (\ref{JackiwHDO}) or (\ref{H10}), can form the Poincar\'{e} algebra together with the momentum and angular momentum generators~\cite{Weinberg}. The Poincar\'{e} algebra ensures the full symmetry of special relativity, which includes translation and rotation symmetries for both time-like and spatial-like directions. In other words, although the theory singles out a particular time parameter for use through the foliation of spacetime, the Poincar\'{e} algebra guarantees that the resulting dynamical evolution is fully relativistic. This is because satisfying this algebra guarantees that one can construct a Poincar\'{e} covariant stress-energy tensor for the field dynamics. 

Explicitly, the Poincar\'{e} algebra consists the following expressions in terms of commutation relations~\cite{Weinberg, Ipek2021} among the Hamiltonian operator $\hat{H}$, the momentum operators $\hat{P}_i$, the angular momentum $\hat{J}_i$, and the Lorentz boost $\hat{K}_i$ $\{i,j,k=1,2,3\}$.
\begin{subequations}
    \label{PoincareAlgebra}
\begin{align}
    [\hat{P}_i, \hat{P}_j] &= 0 \\
    [\hat{P}_i, \hat{H}] &= 0 \\
    [\hat{J}_i, \hat{P}_j] &= i\epsilon_{ijk}\hat{P}_k\\
    [\hat{J}_i, \hat{J}_j] &= i\epsilon_{ijk}\hat{J}_k \\
    [\hat{J}_i, \hat{H}] &= 0 \\
    [\hat{K}_i, \hat{H}] &= i\hat{P}_i \\
    [\hat{K}_i, \hat{P}_j] &= -i\delta_{ij}\hat{H} \\
    [\hat{K}_i, \hat{J}_j] &= -i\epsilon_{ijk}\hat{K}_k \\
    [\hat{K}_i, \hat{K}_j] &= -i\epsilon_{ijk}\hat{J}_k.
\end{align}
\end{subequations}
We wish to define the operators $\hat{P}_i, \hat{J}_i, \hat{K}_i$ properly, which, together with $\hat{H}$ derived in (\ref{JackiwHDO}), can satisfy these commutation relations. 

First of all, we define the momentum operator $\hat{P}_i$ as
\begin{equation}
    \label{momentumOp}
    \hat{P}_i = \int d^3x[(\hat{p}_iu^\dagger)\frac{\delta}{\delta u^\dagger} + (\hat{p}_iu)\frac{\delta}{\delta u}].
\end{equation}
where $\hat{p}_i=-i\partial_i$. Then we have
\begin{equation}
    \label{momentumOp}
    \begin{split} 
    [\hat{P}_i, u^\dagger(y)u(y)] &= \int d^3x[(\hat{p}_iu^\dagger(x))\frac{\delta}{\delta u^\dagger(x)}(u^\dagger(y)u(y))\\
    &+ (\hat{p}_iu(x))\frac{\delta}{\delta u(x)}(u^\dagger(y)u(y))]\\
    &= (\hat{p}_iu^\dagger(y))u(y) - (\hat{p}_iu(y))u^\dagger(y)\\
    &= \hat{p}_i (u^\dagger(y)u(y)).
    \end{split}
\end{equation}
The angular momentum operator is defined in a similar way,
\begin{equation}
    \label{angularOp}
    \hat{J}_i = \int d^3x\epsilon_{ijk}x^j[(\hat{p}^ku^\dagger)\frac{\delta}{\delta u^\dagger} + (\hat{p}^ku)\frac{\delta}{\delta u}].
\end{equation}
Finally, the Lorentz boost is defined as~\cite{Ipek2021}
\begin{equation}
    \label{boost}
    \hat{K}_i = \int d^3x x_i\hat{\mathcal{H}} - t\hat{P}_i,
\end{equation}
where $\hat{\mathcal{H}}$ is the Hamiltonian density operator defined from $\hat{H}=\int d^3x\hat{\mathcal{H}}$. 

With these definitions of $\hat{P}_i, \hat{J}_i, \hat{K}_i$, we show in Appendix \ref{appendix:Poincare} that the Hamiltonian operator (\ref{JackiwHDO}), derived from the extended stationary action principle, satisfies the Poincar\'{e} algebra. Thus, the Schr\"{o}dinger equation \eqref{JackiwSE} meets the symmetry requirements of special relativity. From the proofs in Appendix \ref{appendix:Poincare}, it is clear that the Hamiltonian operator in (\ref{H10}) can also form the Poincar\'{e} algebra with $\hat{P}_i, \hat{J}_i, \hat{K}_i$. This step completes the procedure for quantization of fermionic fields without interactions with other fields.

\section{Recovering Results from Canonical Quantization Approach}
\label{sec:comparison}
To confirm that the quantization framework proposed in this study is consistent with traditional quantization approaches, we will calculate the ground state energy of the vacuum and the probability of particle formation of the fermionic field in a constant external field. The results will be compared with those calculated using the traditional canonical quantization approach. We will show that these results are indeed in agreement. This section closely follows the pioneer work in \cite{Kiefer}, but generalizes the solution with the parameter $\lambda$.

\subsection{Vacuum Energy}
Denote the eigen states of the first quantized Hamiltonian $h$ as $\psi_n$
\begin{equation}
 \label{eigen}   
 h\psi_n = E_n\psi_n, 
\end{equation}
with the completeness relation and the orthogonal condition, respectively, as
\begin{align}
    \sum_n\psi_n(x)\psi_n^\dagger(y) &= \delta(x-y), \\
    \int d^3x\psi_n(x)\psi^\dagger_m(x)&=\delta_{nm}.
\end{align}
We expand $u$ and $u^\dagger$ in terms of these eigen states,
\begin{equation}
    u(x) = \sum_nu_n\psi_n(x), \text{   }u^\dagger(x) = \sum_nu_n^\dagger\psi^\dagger_n(x).
\end{equation}
To ensure $\delta u(x)/\delta u(y)=\delta(x-y)$, we must have
\begin{equation}
    \frac{\delta}{\delta u(x)} = \sum_n\psi^\dagger_n(x)\frac{\delta}{\delta u_n}
\end{equation}
Substituting these identities into \eqref{generalHO2}, we obtain
\begin{equation}
    \label{generalHO3}
    \hat{H} = \frac{\lambda}{4}\sum_n E_n(u_n^\dagger + \frac{2}{\lambda}\frac{\delta }{\delta u_n})(u_n+\frac{2}{\lambda}\frac{\delta }{\delta u^\dagger_n}).
\end{equation}
We can define the particle creation and annihilation operators as
\begin{align}
\label{creation}
    \hat{a}_\alpha &= \frac{\sqrt{\lambda}}{2}(u_\alpha+\frac{2}{\lambda}\frac{\delta }{\delta u^\dagger_\alpha}), \\
    \hat{a}^\dagger_\beta &= \frac{\sqrt{\lambda}}{2}(u^\dagger_\beta + \frac{2}{\lambda}\frac{\delta }{\delta u_\beta}).
\end{align}
where $\alpha, \beta$ are spinor indices. One can verify that they satisfy the anticommutation relation
\begin{equation}
    \{\hat{a}_\alpha,  \hat{a}^\dagger_\beta\} = \delta_{\alpha\beta}\delta(x-y).
\end{equation}
Then, the Hamiltonian operator can be expressed as 
\begin{equation}
    \label{generalHO4}
    \hat{H} = \sum_nE_n\hat{a}^\dagger_n\hat{a}_n.
\end{equation}
Note the spinor indices are suppressed here. A typical choice of the ground state is a Gaussian state~\cite{Kiefer}
\begin{equation}
\label{groundState}
    \Psi_0[u, u^\dagger] = \exp\{\int d^3xd^3y( u^\dagger\Omega u)\},
\end{equation}
where $\Omega$ is expanded as
\begin{equation}
\label{covariance4}
    \Omega(x,y) = \sum_{n,m}\Omega_{nm}\psi_n(x)\psi^\dagger_m(y).
\end{equation}
Using \eqref{generalHO3}, we find that
\begin{equation}
\begin{split}
    \hat{H}\Psi_0 &=\frac{1}{2}\{\sum_n E_n+\frac{2}{\lambda}\sum_nE_n\Omega_{nn} \\
    &+\frac{\lambda}{2}\sum_nE_n[u^\dagger_nu_n-\frac{4}{\lambda^2}(\sum_{i,j}u^\dagger_i\Omega_{in}\Omega_{nj}u_j)]\\
    &+\sum_nE_n[\sum_ju^\dagger_n\Omega_{nj}u_j - \sum_iu^\dagger_i\Omega_{in}u_n]\}\Psi_0.
\end{split}
\end{equation}
Since $\mathrm{Tr}(h)=0$, the first term vanishes. The energy of the ground state $E_0$ of the stationary Schr\"{o}dinger equation $\hat{H}\Psi_0 = E_0\Psi_0$ should not depend on $(u_n, u^\dagger_n)$. Thus, the third and fourth terms must vanish as well, which can be satisfied if
\begin{equation}
    \Omega_{nm} = \pm\frac{\lambda}{2}\delta_{nm}. 
\end{equation}
This leaves
\begin{equation}
\label{gE}
    E_0 =  \frac{1}{\lambda}\sum_nE_n\Omega_{nn}.
\end{equation}
We also demand that by applying the annihilation and creation operator to the ground state,
\begin{equation}
    \hat{a}^\dagger_n\hat{a}_n\Psi_0 = (\frac{1}{2}+\frac{1}{\lambda}\Omega_{nn})\Psi_0,
\end{equation}
the resulting state should be null for positive energy. That is,
\begin{equation}
    \hat{a}^\dagger_n\hat{a}_n\Psi_0 = \left\{ \begin{array}{lllll} 0 &\mbox{if}&\Omega_{nn}=-\lambda/2,&\text{for}&E_n > 0 \\ \Psi_0 &\text{if}&\Omega_{nn}=\lambda/2, &\text{for}&E_n < 0 \end{array}\right.
\end{equation}
Then, the ground state energy \eqref{gE} can be written without ambiguity the vacuum energy in momentum space,
\begin{equation}
\label{gE2}
    E_0 =  -\frac{1}{2}\sum_n|E_n| = -\frac{1}{2}\frac{V}{(2\pi)^3}\int d^3p\sqrt{p^2+m^2}.
\end{equation}
Eq. \eqref{gE2} is the same result derived in \cite{Kiefer}, and in agreement with the result using the canonical quantization approach \cite{Zee}. It does not depend on the parameter $\lambda$.

The covariance $\Omega$ in \eqref{covariance4} can be further simplified as
\begin{equation}
\label{covariance2}
    \Omega(x,y) = \frac{\lambda}{2} (\sum_{E_n<0}\psi_n(x)\psi^\dagger_n(y) - \sum_{E_n>0}\psi_n(x)\psi^\dagger_n(y)).
\end{equation}
It is convenient to define two projectors
\begin{equation}
    P_- = \sum_{E_n<0}\psi_n(x)\psi^\dagger_n(y), \text{ } P_+=\sum_{E_n>0}\psi_n(x)\psi^\dagger_n(y).
\end{equation}
Given the orthogonal properties of $\{\psi_n\}$, one has
\begin{equation}
    P_-+P_+ = 1, \text{ } (P_-)^2 = P_-,  \text{ } (P_+)^2 = P_+,\text{ }  P_+P_-=0.
\end{equation}
Thus, 
\begin{equation}
\label{covariance3}
    \Omega = \frac{\lambda}{2}(P_- - P_+), \text{ } \Omega^2 = \frac{\lambda^2}{4}.
\end{equation}

With the definitions of the creation and annihilation operators \eqref{creation}, the Hamiltonian operator in \eqref{H10} cannot be written in the well-known format \eqref{generalHO4}. It is not clear how the creation and annihilation operators should be defined such that the Hamiltonian operator in \eqref{H10} can be written in the format \eqref{generalHO4}. This shows an advantage of the Floreanini-Jackiw representation, which is well adopted in the research literature for the functional representation of Schr\"{o}dinger representation of fermionic fields. The form of Hamiltonian operator in \eqref{H10} is less preferred and is used only for heuristic purposes.

\subsection{Solution of the Time-dependent Schr\"{o}dinger Equation}
We again closely follow the technique in \cite{Kiefer} to derive the solution of the time-dependent Schr\"{o}dinger equation \eqref{generalSE}, but generalize the solution with the parameter $\lambda$. Again, assuming that the general solution of \eqref{generalSE} is Gaussian, we can write
\begin{equation}
\label{timeEvolState}
    \Psi = N(t)\exp(u^\dagger\Omega(t)u),
\end{equation}
where $N$ and $\Omega$ are time-dependent. It can be considered as the time evolution of the vacuum state \eqref{groundState}. Inserting \eqref{timeEvolState} into \eqref{generalSE}, we obtain
\begin{align}
    \label{factorNeq}
    i\frac{d\ln N}{dt} &= \frac{1}{\lambda}\mathrm{Tr}(h\Omega) \\
    \label{covariance}
    i\dot{\Omega} &= \frac{\lambda}{4}(1 - \frac{2}{\lambda}\Omega)h(1 + \frac{2}{\lambda}\Omega).
\end{align}
Eq.\eqref{covariance} is solved by introducing an operator $Q(t)$ which satisfies
\begin{equation}
\label{DiracEq}
    i\dot{Q} = hQ,
\end{equation}
and a time-independent operator $C$. Then,
\begin{equation}
    \label{solCovariance}
    \Omega(t) = \frac{\lambda}{2}(Q(t) - C)(Q(t)+C)^{-1}.
\end{equation}
It is intuitive to demand that $\Omega$ takes the ``free solution" \eqref{covariance3} in the asymptotic past. That is, $\Omega\to\Omega_0 =(\lambda/2)( P_- - P_+)$ when $t\to -\infty$. This is equivalent to choosing $C=P_+$ and $Q(t)\to P_-$ when $t\to -\infty$. Then, the solution to $\Omega(t)$ is reduced to solving equation \eqref{DiracEq} with the first quantized Hamiltonian. 

Once $\Omega(t)$ is obtained, the factor $N(t)$ can be determined from \eqref{factorNeq}
\begin{equation}
    N(t) = N_0\exp{(-\frac{i}{\lambda}\int^t \mathrm{Tr}(h\Omega)ds)},
\end{equation}
where $N_0$ can be fixed by the normalization condition in \eqref{Normalization}, and one finds
\begin{equation}
\label{factorN}
    N(t) = (\det(1+\Omega^\dagger\Omega))^{-1/2}\exp\{-\frac{i}{\lambda}\int^t Re(\mathrm{Tr}(h\Omega))ds\}.
\end{equation}
With \eqref{timeEvolState} and \eqref{factorN}, we can calculate the absolute square of the matrix element of two Gaussian states, $\Psi_1$ and $\Psi_2$, as the following expression,
\begin{equation}
    \label{matrixElem}
    |\langle\Psi_1|\Psi_2\rangle|^2 = \det \frac{(1+\Omega_1^\dagger\Omega_2)(1+\Omega_2^\dagger\Omega_1)}{(1+\Omega_1^\dagger\Omega_1)(1+\Omega_2^\dagger\Omega_2)}.
\end{equation}
This is the basic expression for calculating the probability of particle creation.

\subsection{Probability of Particle Creation}
\label{subsec:PC}
Suppose that we take $\Psi_1$ as the vacuum state in the asymptotic past and $\Omega_1\to\Omega_0$ in \eqref{covariance3}, and $\Psi_2$ is the state of time evolving from the vacuum state according to \eqref{generalSE}, where the covariance $\Omega_2(t)$ evolves according to \eqref{covariance} and the Hamiltonian $h$ contains an external electromagnetic field, then expression \eqref{matrixElem} is interpreted as the probability of particle creation. Here we assume that $\Psi_2$ will take the vacuum state again in the asymptotic future. This corresponds to a physically reasonable assumption that the external field is switched on somewhere in the past and switched off again somewhere in the future. We will first compute \eqref{matrixElem} when $\Omega_1=\Omega_0$ and $\Omega_2$ as the general solution to \eqref{covariance}, then take the solution of $\Omega_2$ at the asymptotic limit $t\to \infty$. We will remove the subscript of $\Omega_2$ in the remainder of this section. 

Since $\Omega_0=\Omega_0^\dagger$ and $\Omega_0^2 = \lambda^2/4$, \eqref{matrixElem} is rewritten as
\begin{equation}
    \label{matrixElem2}
    |\langle\Psi_1|\Psi_2\rangle|^2 = \det \frac{(1+\Omega_0\Omega(t))(1+\Omega^\dagger(t)\Omega_0)}{(1+\lambda^2/4)(1+\Omega^\dagger(t)\Omega(t))},
\end{equation}
where $\Omega(t)$ takes the form of \eqref{solCovariance} with $C=P_+$,
\begin{equation}
    \label{solCovariance}
    \Omega(t) = \frac{\lambda}{2}(Q(t) - P_+)(Q(t)+P_+)^{-1}.
\end{equation}
$Q(t)$ is determined by solving \eqref{DiracEq} and, in general, can be expanded as~\cite{Kiefer}
\begin{equation}
    Q(t) = \sum_n\chi_n(t)\chi_n^\dagger,
\end{equation}
where $\chi_n$ is an eigenfunction of the first quantized Hamiltonian $h$ with negative frequency, and $\chi_n(t)$ denotes the solution of \eqref{DiracEq} that satisfies $\lim_{t\to -\infty}\chi_n(t) = \chi_n$. Further expanding $\chi_n(t)$ as
\begin{equation}
    \chi_n(t) = \alpha_{nm}(t)\chi_m + \beta_{nm}(t)\psi_m,
\end{equation}
where $\psi_m$ is a positive frequency eigenfunction of $h$, and $\alpha, \beta$ are the time-dependent Bogolubov coefficients that satisfy the normal condition $\alpha^\dagger\alpha+\beta^\dagger\beta =1$. Following calculations similar to \cite{Kiefer}, we find that
\begin{equation}
\label{OmegaB}
    \Omega(t) = \Omega_0 + \lambda B,
\end{equation}
where $B$ is an operator defined as
\begin{equation}
\label{opB}
    B(x,y) = \sum_{n,s,t}\psi_n(x)\alpha^{-1}_{st}\beta_{tn}\chi_s^\dagger(y).
\end{equation}
The operator $B$ is considered a nilpotent operator, which maps negative energy eigenfunctions to positive ones and annihilates positive energy eigenfunctions. 

From \eqref{OmegaB} and the fact that $\Omega_0=\lambda^2/4$, one can derive the following expressions
\begin{align}
    \Omega_0\Omega(t) &= \frac{\lambda^2}{4}(1-2B),\\
    \Omega^\dagger(t)\Omega_0 &= \frac{\lambda^2}{4}(1-2B^\dagger),\\
    \Omega^\dagger(t)\Omega(t) &= \frac{\lambda^2}{4}(1-2B-2B^\dagger+4B^\dagger B).
\end{align}
Substituting them into \eqref{matrixElem2}, we have
\begin{equation}
    \label{matrixElem3}
    |\langle\Psi_1|\Psi_2\rangle|^2 = \det \frac{(1+\frac{\lambda^2}{4}(1-2B))(1+\frac{\lambda^2}{4}(1-2B^\dagger))}{(1+\frac{\lambda^2}{4})(1+\frac{\lambda^2}{4}(1-2B-2B^\dagger+4B^\dagger B))}.
\end{equation}
Written in the basis $(\psi,\chi)^T$, operator $B$, defined in \eqref{opB}, and its adjoint $B^\dagger$, read
\begin{equation}
    B = \left ( \begin{array}{cc} 0 & \alpha^{-1}\beta \\ 0& 0 \end{array}\right ), \text{  } 
    B^\dagger = \left ( \begin{array}{cc} 0 & 0 \\ (\alpha^{-1}\beta)^\dagger& 0 \end{array}\right ).
\end{equation}
For simpler notation, denote $\nu=(1+\lambda^2/4)$. One can verify that
\begin{align*}
    &\det (1+\frac{\lambda^2}{4}(1-2B)) = \det (\nu^2)\\
    &\det (1+\frac{\lambda^2}{4}(1-2B^\dagger)) = \det (\nu^2)\\
    &\det (1+\frac{\lambda^2}{4}(1-2B-2B^\dagger+4B^\dagger B)) \\ &=\det(\nu^2)(\nu^2+\lambda^2\alpha^{-1}\beta(\alpha^{-1}\beta)^\dagger).
\end{align*}
Substituting them into \eqref{matrixElem3}, we find that
\begin{equation}
    \label{matrixElem4}
    |\langle\Psi_1|\Psi_2\rangle|^2 = \det (1+\kappa\alpha^{-1}\beta(\alpha^{-1}\beta)^\dagger)^{-1},
\end{equation}
where $\kappa = (\lambda/\nu)^2=\lambda^2/(1+\lambda^2/4)^2$. Given the property $\alpha^\dagger\alpha+\beta^\dagger\beta =1$, \eqref{matrixElem4} can be further simplified as
\begin{equation}
    \label{matrixElem5}
    |\langle\Psi_1|\Psi_2\rangle|^2 = \det (1+\kappa\beta^\dagger(1-\beta\beta^\dagger)^{-1}\beta)^{-1}.
\end{equation}
Note that when $\lambda=2$, we have $\kappa=1$, then \eqref{matrixElem5} is reduced to the same result as in \cite{Kiefer}.

Eq.\eqref{matrixElem5} shows that the calculation of the particle creation probability is reduced to finding the Bogolubov coefficients $\beta$, which are completely determined by \eqref{DiracEq}. For a charged fermionic field under the influence of a constant external electric field $\vec{E}=E\vec{e}_z$, it turns out that \cite{Kiefer}
\begin{equation}
    \label{beta}
    \beta = e^{-\frac{\pi\xi}{2}},
\end{equation}
where
\begin{equation}
    \xi = \frac{p_x^2+p_y^2+m^2}{|eE|}.
\end{equation}
These allow us to evaluate \eqref{matrixElem5} as
\begin{equation}
    \label{matrixElem6}
    \begin{split}
         |\langle\Psi_1|\Psi_2\rangle|^2 &= \det (1-\frac{\kappa e^{-\pi\xi}}{1+(\kappa-1)e^{-\pi\xi}}) \\ &=\exp \mathrm{Tr}(\ln(1-\frac{\kappa e^{-\pi\xi}}{1+(\kappa-1)e^{-\pi\xi}}))\\ &= \exp(-\mathrm{Tr}\sum_n \frac{1}{n}(\frac{\kappa e^{-\pi\xi}}{1+(\kappa-1)e^{-\pi\xi}})^n).
    \end{split}
\end{equation}
The trace operator reads \cite{Kiefer}
\begin{equation}
    \mathrm{Tr} \to \frac{2V}{(2\pi)^3}\int dp_xdp_ydp_z=\frac{2eEVT}{(2\pi)^3}\int dp_xdp_y,
\end{equation}
where $T$ is the time duration of the influence of the external electric field, which expands from the far past (``in - region") to the far future (``out - region"). Thus,
\begin{equation}
    \label{matrixElem7}
    \begin{split}
    &|\langle\Psi_1|\Psi_2\rangle|^2 = \\ &\exp(-\frac{2eEVT}{(2\pi)^3}\sum_n\frac{1}{n}\int dp_xdp_y(\frac{\kappa e^{-\pi\xi}}{1+(\kappa-1)e^{-\pi\xi}})^n).
    \end{split}
\end{equation}
It is not possible to carry out the integrals over $p_x$ and $p_y$ analytically. However, when $\kappa=1$, that is, when $\lambda=2$, the integrand in \eqref{matrixElem7} becomes Gaussian and the integrals can be carried out,
\begin{equation}
    \label{matrixElem8}
    \begin{split}
    |\langle\Psi_1|\Psi_2\rangle|^2 &= \exp(-\frac{2eEVT}{(2\pi)^3}\sum_n\frac{1}{n}\int dp_xdp_ye^{-n\pi\xi})\\
    &= \exp(-\frac{2(eE)^2VT}{(2\pi)^3}\sum_n\frac{1}{n^2}e^{-\frac{n\pi m^2}{eE}}).
    \end{split}
\end{equation}
This agrees exactly with the classical result of Schwinger \cite{Schwinger} using the canonical quantization approach.

\section{Field Interactions}
\label{sec:interactions}
In this section, we apply the quantization framework to the Lagrangian that includes interaction with other fields. Specifically, we will quantize the fermionic fields that are coupling with Abelian electromagnetic fields, or non-Abelian gauge fields, or interacting with the fermionic field itself. The last case will lead to a nonlinear functional Schr\"{o}dinger equation.

\subsection{Interaction with Electromagnetic Field}
Adding the interaction term between the ferminonic field and the electromagnetic vector field $\mathbf{A}$ into the Lagrangian \eqref{LD2} amounts to promoting the regular derivative operator to a covariant derivative, 
\begin{equation}
\label{LDE}
    \mathcal{L} = \frac{i}{2}\bar{\psi}\gamma^\mu D_\mu\psi - \frac{i}{2}(D_\mu\bar{\psi})\gamma^\mu\psi - m\bar{\psi}\psi,
\end{equation}
where the covariant derivative is defined as $D_\mu \psi= (\partial_\mu\ +ieA_\mu)\psi$ and $D_\mu \bar{\psi}= (\partial_\mu -ieA_\mu)\bar{\psi}$. Expanding $D_\mu$ in \eqref{LDE},
\begin{equation}
\label{LDE2}
    \mathcal{L} = \mathcal{L}_0 - e\bar{\psi}\gamma^\mu A_\mu\psi,
\end{equation}
where $\mathcal{L}_0$ is the Lagrangian density of the free fermionic field \eqref{LD2}. This extra term is quadratic in the sense that it is in the form of $\bar{\psi}\Omega\psi$ where the matrix is $\Omega=\gamma^\mu A_\mu$. 
From this Lagrangian density, the momentum conjugates to the field variables $\psi$ and $\psi^\dagger$ are still given by \eqref{momentum1} and \eqref{momentum2}. Choosing the gauge condition $A_0=0$, the Hamiltonian becomes
\begin{equation}
    \label{H7}
    \begin{split}
    H =\int dxdy\psi^\dagger(x) h'(x,y)\psi(y),
    \end{split}
\end{equation}
where we have suppressed the superscript in $d^3xd^3y$ for simpler notation, and define
\begin{equation}
    \label{hprime}
    \begin{split}
    h'(x,y) &= -i\gamma^0\gamma^i\partial_i\delta(x-y) + \gamma^0(m+e\gamma^i A_i)\delta(x-y)\\
    &= h(x,y) + e\gamma^0\gamma^i A_i\delta(x-y).       
    \end{split}
\end{equation}
After the extended canonical transformation, the Hamiltonian is similar to \eqref{H3} but with $h$ replaced by $h'$.
\begin{equation}
    \label{HA}
    H = -\frac{4}{\lambda^2}\int dxdy \{\frac{\delta S}{\delta\psi}h'\frac{\delta S}{\delta\psi^\dagger} \}.
\end{equation}
Or, it can be in the more general form
\begin{equation}
    \label{HA2}
    H = \frac{1}{4}\int dxdy (\psi^\dagger + \frac{2i}{\lambda}\frac{\delta S}{\delta\psi})h'(\psi+\frac{2i}{\lambda}\frac{\delta S}{\delta\psi^\dagger}).
\end{equation}
The rest of the quantization procedure is the same as the quantization of free fermionic fields, only with $h$ replaced by $h'$. The resulting Floreanini-Jackiw representation of functional Schr\"{o}dinger equation is
\begin{equation}
    \label{JackiwSEA}
    i\partial_0\Psi = \frac{\lambda}{4}\{\int dxdy (\psi^\dagger + \frac{2}{\lambda}\frac{\delta }{\delta \psi})h'(\psi+\frac{2}{\lambda}\frac{\delta }{\delta \psi^\dagger})\}\Psi.
\end{equation}

\subsection{Interaction with Non-Abelian Gauge Field}

Consider a toy theory of local $SU(2)$ symmetry for two types of fermions, each with mass $m$~\cite{LancasterBook}. The fermionic fields can be written as
\begin{equation}
    \Psi = \left(\begin{array}{c}\psi_1 \\ \psi_2 \end{array}\right), \bar{\Psi}=(\bar{\psi}_1, \bar{\psi}_2)=(\psi_1^\dagger,\psi_2^\dagger)\gamma^0.
\end{equation}
The standard Lagrangian density, similar to \eqref{LD}, is
\begin{equation}
    \label{LD3}
    \mathcal{L} = \bar{\Psi}(i\gamma^\mu\partial_{\mu}-m)\Psi.
\end{equation}
The $SU(2)$ gauge theory is then described by the following Lagrangian density~\cite{LancasterBook}
\begin{align}
    \label{GaugeLD}
    &\mathcal{L}_g = \bar{\Psi}(i\gamma^\mu D_{\mu}-m)\Psi -\frac{1}{4}G_{\mu\nu}\cdot G^{\mu\nu},\mbox{ where}\\
    &D_{\mu} =\partial_{\mu} - \frac{i}{2}g\mathbf{\tau}\cdot \mathbf{W}_{\mu}(x),\\
    &G_{\mu\nu}= \partial_{\mu}\mathbf{W}_{\nu}-\partial_{\nu}\mathbf{W}_{\mu}+g(\mathbf{W}_{\mu}\times \mathbf{W}_{\nu}),
\end{align}
$\tau$ is the Pauli matrices for isospin, and $g$ is the charge of the theory that determines how strong the gauge field $\mathbf{W}_{\mu}$ interacts with $\Psi$. 

However, if we use the form of Lagrangian density similar to \eqref{LD2}, we have
\begin{equation}
    \mathcal{L}' = \frac{i}{2}\bar{\Psi}\gamma^\mu\partial_\mu\Psi - \frac{i}{2}\partial_\mu\bar{\Psi}\gamma^\mu\Psi - m\bar{\Psi}\Psi,
\end{equation}
and the correspondent Lagrangian density for the gauge theory is
\begin{equation}
\label{GaugeLD2}
    \mathcal{L}'_g = \frac{i}{2}\bar{\Psi}\gamma^\mu D_\mu\Psi - \frac{i}{2}(D_\mu\bar{\Psi})\gamma^\mu\Psi - m\bar{\Psi}\Psi -\frac{1}{4}G_{\mu\nu}\cdot G^{\mu\nu}.
\end{equation}
The covariant derivative acting on $\bar{\Psi}$ is defined as
\begin{equation}
    D_\mu\bar{\Psi} = \partial_{\mu}\bar{\Psi} + \frac{i}{2}g\bar{\Psi}(\mathbf{\tau}\cdot \mathbf{W}_{\mu})^\dagger.
\end{equation}
The question here is whether $\mathcal{L}'_g$ is equivalent to $\mathcal{L}_g$. Expanding the covariant derivative in \eqref{GaugeLD}, we have
\begin{equation}
\label{GaugeLD3}
    \mathcal{L}_g = \mathcal{L}+ \frac{g}{2}\bar{\Psi}(\gamma^\mu \mathbf{\tau}\cdot \mathbf{W}_{\mu})\Psi -\frac{1}{4}G_{\mu\nu}\cdot G^{\mu\nu}.
\end{equation}
The second term is the minimal coupling term that describes the interaction between the gauge field $\mathbf{W}_{\mu}$ and the fermionic field $\Psi$. Without considering the quantization of the gauge field itself, we can drop the last term in \eqref{GaugeLD3}, and rewrite it as
\begin{equation}
\label{GaugeLD5}
    \mathcal{L}_g = \bar{\Psi}(i\gamma^\mu\partial_\mu-m+\frac{g}{2}\gamma^\mu \mathbf{\tau}\cdot \mathbf{W}_{\mu})\Psi.
\end{equation}
This is a quadratic form and the quantization procedure is again similar to that presented earlier.

If we use the form of Lagrangian density similar to \eqref{LD2}, we have
\begin{equation}
    \mathcal{L}' = \frac{i}{2}\bar{\Psi}\gamma^\mu\partial_\mu\Psi - \frac{i}{2}\partial_\mu\bar{\Psi}\gamma^\mu\Psi - m\bar{\Psi}\Psi,
\end{equation}
and the correspondent Lagrangian density for the gauge theory is
\begin{equation}
\label{GaugeLD2}
    \mathcal{L}'_g = \frac{i}{2}\bar{\Psi}\gamma^\mu D_\mu\Psi - \frac{i}{2}(D_\mu\bar{\Psi})\gamma^\mu\Psi - m\bar{\Psi}\Psi -\frac{1}{4}G_{\mu\nu}\cdot G^{\mu\nu}.
\end{equation}
The covariant derivative acting on $\bar{\Psi}$ is defined as
\begin{equation}
    D_\mu\bar{\Psi} = \partial_{\mu}\bar{\Psi} + \frac{i}{2}g\bar{\Psi}(\mathbf{\tau}\cdot \mathbf{W}_{\mu}).
\end{equation}
Expanding the covariant derivative in \eqref{GaugeLD2}, we find
\begin{equation}
\label{GaugeLD4}
    \mathcal{L}'_g = \mathcal{L}'+ \frac{g}{2}\bar{\Psi}(\mathbf{\tau}\cdot \mathbf{W}_{\mu})^\dagger\gamma^\mu\Psi -\frac{1}{4}G_{\mu\nu}\cdot G^{\mu\nu},
\end{equation}
which is in a form similar to \eqref{GaugeLD3}, as desired.

\subsection{Interaction Between Fermions}
\label{sec:NLSE}
Now we consider a more complicated Lagrangian for the Fermi's theory of weak interaction between fermions~\cite{LancasterBook}
\begin{equation}
    \label{FermiLD}
        \mathcal{L} = \frac{i}{2}\bar{\psi}\gamma^\mu \partial_\mu\psi - \frac{i}{2}(\partial_\mu\bar{\psi})\gamma^\mu\psi - m\bar{\psi}\psi + G(\bar{\psi}\psi)^2,
\end{equation}
where $G$ is a coupling constant that determines the strength of interaction between the fermions. Clearly, the interaction term is no longer quadratic. In fact, the theory with such a Lagrangian density is non-renormalizable~\cite{LancasterBook}. It would be interesting to see whether our quantization framework can be applied for such a Lagrangian. 

The momentum conjugates to the field variables $\psi$ and $\psi^\dagger$ are still given by \eqref{momentum1} and \eqref{momentum2}. The Hamiltonian is calculated similarly to \eqref{H2} as
\begin{equation}
    \label{H8}
    \begin{split}
    H =\int dxdy\{(\psi^\dagger h\psi) - (\psi^\dagger g\psi)^2\},
    \end{split}
\end{equation}
where we denote $g=\gamma^0\sqrt{G}\delta(x-y)$. 

Step \textbf{II}. Performing the canonical transformation. Eqs. \eqref{Conjugate} to \eqref{action2} are still valid, and the Hamiltonian becomes
\begin{equation}
    \label{H8}
    \begin{split}
    H =-\int dxdy\{\frac{4}{\lambda^2}(\frac{\delta S}{\delta\psi}h\frac{\delta S}{\delta\psi^\dagger}) + \frac{16}{\lambda^4}(\frac{\delta S}{\delta\psi}g\frac{\delta S}{\delta\psi^\dagger})^2\}.
    \end{split}
\end{equation}
The action functional for the field ensemble is 
\begin{equation}
    \label{ensembleAction2}
    \begin{split}
    S_c =& \int dt\mathcal{D}\psi^\dagger\mathcal{D}\psi \{\rho (-\frac{\partial S}{\partial t} \\
    &+\int dxdy[\frac{4}{\lambda}(\frac{\delta S}{\delta\psi}h\frac{\delta S}{\delta\psi^\dagger}) + \frac{16}{\lambda^3}(\frac{\delta S}{\delta\psi}g\frac{\delta S}{\delta\psi^\dagger})^2]\}.
    \end{split}
\end{equation}

Step \textbf{III}. Following similar derivations in Section IV, we can obtain the functional probability density for field fluctuations in an infinitesimal time step $\Delta t$ as
\begin{equation}
\label{transP3}
    p[\omega, \omega^\dagger] = \frac{1}{Z}\exp\{2\Delta t\int dxdy[(\omega^\dagger h\omega)-(\omega^\dagger g\omega)^2]\}.
\end{equation}
This is no longer a Gaussian functional. Calculating the expectation value $\langle\omega_\alpha\omega^\dagger_\beta\rangle$ is not easy given the definition of inner product in Appendix \ref{appendix:exp}. To proceed further, we can assume that, in the infinitesimal time step, the contribution from the interaction to the field fluctuations can be ignored. This means that the second term in the exponential of \eqref{transP3} is ignored assuming that $g$ is sufficiently small. Consequently, the probability density \eqref{transP3} is reduced to \eqref{transP}, and $\langle\omega_\alpha\omega^\dagger_\beta\rangle$ is still given by \eqref{expectation}.

Step \textbf{IV} The information metrics of field fluctuations for a period of time, $I_f$, is still defined in \eqref{TDivergence} using the Tsallis divergence, which is further simplified to \eqref{Tsallis} given \eqref{expectation}. Together with \eqref{ensembleAction2}, the total action functional is
\begin{equation}
    \label{ensembleAction3}
    \begin{split}
    S_t =& \int dt\mathcal{D}\psi^\dagger\mathcal{D}\psi \{\rho [-\frac{\partial S}{\partial t} 
    +\int dxdy[\frac{4}{\lambda}(\frac{\delta S}{\delta\psi}h\frac{\delta S}{\delta\psi^\dagger})\\
    &+ \frac{16}{\lambda^3}\Theta^2 ]+\frac{\alpha}{2\rho}(\frac{\delta \rho}{\delta\psi}h\frac{\delta \rho}{\delta\psi^\dagger})\}.
    \end{split}
\end{equation}
where $\Theta$ is a functional introduced to simplify notation
\begin{equation}
    \Theta = \frac{\delta S}{\delta\psi}g\frac{\delta S}{\delta\psi^\dagger}.
\end{equation}

Step \textbf{V} Variation of \eqref{ensembleAction3} over $\rho$ gives the quantum version of Hamilton-Jacobi equation
\begin{equation}
\label{QHJ3}
\begin{split}
    \frac{\partial S}{\partial t} =& \int dxdy\{\frac{\lambda}{4}\frac{\delta S}{\delta\psi}h\frac{\delta S}{\delta\psi^\dagger} + \frac{16}{\lambda^3}\Theta^2 -\frac{2\alpha}{R}\frac{\delta R}{\delta\psi}h\frac{\delta R}{\delta\psi^\dagger}\},
    \end{split}
\end{equation}

The variation of \eqref{ensembleAction3} over $S$ is more complicated and results in
\begin{equation}
\label{ContEq2}
    \begin{split}
        \frac{\partial\rho}{\partial t} &= \frac{4}{\lambda}\int dxdy \{\frac{\delta \rho}{\delta\psi}h\frac{\delta S}{\delta\psi^\dagger}+ \frac{\delta S}{\delta\psi}h\frac{\delta \rho}{\delta\psi^\dagger} + 2\rho\frac{\delta }{\delta\psi}h\frac{\delta S}{\delta\psi^\dagger}\}\\
        &+\frac{32}{\lambda^3}\int dxdy \{(\frac{\delta \rho}{\delta\psi}g\frac{\delta S}{\delta\psi^\dagger}+ \frac{\delta S}{\delta\psi}g\frac{\delta \rho}{\delta\psi^\dagger}+2\rho\frac{\delta }{\delta\psi}g\frac{\delta S}{\delta\psi^\dagger})\Theta \\
        &+\rho(\frac{\delta \Theta}{\delta\psi}g\frac{\delta S}{\delta\psi^\dagger}+ \frac{\delta S}{\delta\psi}g\frac{\delta \Theta}{\delta\psi^\dagger})\}.
    \end{split}
\end{equation}
Using the definition of $\Psi$ in \eqref{wavefunctional}, and choosing $\alpha=2/\lambda$, we combine \eqref{QHJ3} and \eqref{ContEq2} into a single equation with functional derivative,
\begin{equation}
    \label{NLSE}
        i\partial_0 \Psi = \frac{4}{\lambda}\int dxdy (\frac{\delta }{\delta\psi}h\frac{\delta}{\delta\psi^\dagger})\Psi + \Lambda\Psi,
\end{equation}
where the functional
\begin{equation}
\label{Lamda}
\begin{split}
        \Lambda =& \frac{16}{\lambda^3}\int dxdy \{(\frac{\delta \rho}{\delta\psi}g\frac{\delta S}{\delta\psi^\dagger}+ \frac{\delta S}{\delta\psi}g\frac{\delta \rho}{\delta\psi^\dagger}+2\rho\frac{\delta }{\delta\psi}g\frac{\delta S}{\delta\psi^\dagger})\Theta \\
        &+\rho(\frac{\delta \Theta}{\delta\psi}g\frac{\delta S}{\delta\psi^\dagger}+ \frac{\delta S}{\delta\psi}g\frac{\delta \Theta}{\delta\psi^\dagger}) - \Theta^2\}.
\end{split}
\end{equation}
Taking the complex conjugate of $\Psi$ in \eqref{wavefunctional}, and denoting $\bar{\Psi}=Re^{-iS}$, we have
\begin{equation}
\label{rhoS}
    \rho=\bar{\Psi}\Psi, S=\frac{i}{2}(\ln\bar{\Psi}-\ln{\Psi}).
\end{equation}
Substituting $\rho$ and $S$ in \eqref{Lamda} with \eqref{rhoS}, and expressing $\Lambda$ in terms of $\Psi$ and $\bar{\Psi}$, the resulting expression is non-trivial and cannot be simplified as an operator acting on $\Psi$. Instead, $\Lambda$ is a functional of $\Psi$ and $\bar{\Psi}$, so that
\begin{equation}
    \label{NLSE2}
        i\partial_0 \Psi = \hat{H}_0\Psi + \Lambda(\Psi, \bar{\Psi})\Psi,
\end{equation}
where $\hat{H}_0$ is Hamiltonian operator for the free fermionic fields as defined in \eqref{H10}. On the other hand, if we follow the standard canonical quantization procedure and promote $\psi\to \frac{2}{\sqrt\lambda}\frac{\delta}{\delta \psi^\dagger}$ and $\psi^\dagger\to \frac{2}{\sqrt\lambda}\frac{\delta}{\delta \psi}$ in \eqref{H8}, we obtain a linear Schr\"{o}dinger equation
\begin{equation}
    \label{SE5}
     i\partial_0 \Psi = \hat{H}_0\Psi - \frac{16}{\lambda^2}\int dxdy(\frac{\delta }{\delta\psi}g\frac{\delta }{\delta\psi^\dagger})^2\Psi.
\end{equation}
Detailed calculation shows that the second term in \eqref{SE5} is different from the second term in \eqref{NLSE2}. Eq. \eqref{NLSE2} is a non-linear equation of $\Psi$ with functional derivative. In general, there is no guarantee that a linear Schr\"{o}dinger equation always exists for a non-renormalizable quantum field theory~\cite{Symanzik}. The result in \eqref{NLSE2} confirms such an assertion in the case of quantum field theory for Fermion interactions.

\section{Discussion and conclusions} 
\label{sec:discussion}

\subsection{Comparisons with Standard Second Quantization Frameworks}
The two standard second quantization frameworks in quantum field theory, canonical quantization and the path integral formulation, as well as the quantization framework presented in this paper, all originate from the Lagrangian formalism. Among them, the path integral formulation is often considered the most straightforward. However, it implicitly assumes the existence of a linear Schrödinger equation for the wave functional. In fact, the path integral formulation and the linear Schrödinger equation can be derived from each other~\cite{Feynman48, FeynmanBook}. This raises an interesting question: Can the path integral approach be applied to quantize fields described by Lagrangians such as \eqref{FermiLD} because our quantization framework demonstrates that, for such a Lagrangian, the resulting Schrödinger equation is inherently non-linear. 

Both canonical quantization and the quantization framework presented in this paper derive the conjugate momenta from the Lagrangian. However, in canonical quantization, the field variables and their conjugate momenta are promoted to operators as a fundamental postulate, an assumption that can sometimes appear ad hoc, as seen in the Floreanini-Jackiw representation of fermionic fields. In contrast, the quantization framework developed here does not require this operator promotion step. Instead, operators emerge naturally as mathematical tools after the quantization process. Furthermore, standard canonical quantization also assumes the existence of a linear Schrödinger equation in the wave functional representation. But our result shows that for fields governed by Lagrangians such as \eqref{FermiLD}, the Schrödinger equation is non-linear. 

\subsection{Physical Meanings of the Free Parameters}
There are two parameters introduced in the quantization framework presented in this paper, the order of Tsallis divergence $\alpha$, and the canonical transformation parameter $\lambda$. Since the canonical transformation does not alter the underlying physical dynamics, there is no physical significance of the parameter $\lambda$. Instead, it is a parameter that is used to meet the requirement of linearity of the Schr\"{o}dinger equation of the functional wave in conjunction with the parameter $\alpha$. That is, they should satisfy the relation $\alpha \lambda= 1/2$ to ensure the linearity of the Floreanini-Jackiw representation of the Schr\"{o}dinger equation. Thus, fixing $\alpha$ also fixed $\lambda$, or vice versa. 

The parameter $\alpha$, on the other hand, contains physical meanings related to the effect of field fluctuation. The Tsallis divergence quantifies the deviation of the probability distribution due to field fluctuation from the probability distribution without field fluctuation. To help understand the meanings of $\alpha$, we can use a simpler discrete form of the definition of $D_T^\alpha$ in \eqref{TDivergence}
\begin{equation}
    D_T^\alpha = \frac{1}{\alpha -1}(\sum_iq_i(\frac{p_i}{q_i})^\alpha -1),
\end{equation}
where $p_i$ is the probability of $i$-th field configuration without field fluctuation and $q_i$ is the probability of the same field configuration but with field fluctuation. When $\alpha\to 1$, there is no weight on each $r_i = p_i/q_i$, and the Tsallis divergence converges to the Kullback-Leibler divergence. However, for $\alpha\ne 1$, there is a weight on each $r_i$. When $\alpha\to 0$, $r_i^\alpha \to 1$ regardless of the value of $r_i$; On the other hand, when $\alpha \gg 1$, $r_i^\alpha \to 0$ for $0 < r_i < 1$ and $r_i^\alpha \to \infty$ for $ r_i > 1$. Thus, when $\alpha < 1$, the deviation of the probability distribution due to field fluctuation is deemphasized, while when $\alpha > 1$, the deviation is enhanced. $\alpha$ serves as a parameter to regulate the degree to which the fluctuation of the field causes the probability distribution to deviate from its original distribution in a fermionic field.

There is only one free parameter given the relation $\alpha \lambda= 1/2$. Although we keep $\lambda$ instead of $\alpha$ in the Floreanini-Jackiw representation of the Schr\"{o}dinger equation due to mathematical convenience, it is actually more physically meaningful to replace $\lambda$ with $\alpha$. The theory presented here does not predict the exact value of $\alpha$. Instead, it is determined by comparing it with the experimental verifiable physical value. For example, comparing it with the theoretical calculation of the probability of particle creation, as shown in Section \ref{subsec:PC}, we determine that $\alpha = 1/4$ (and therefore $\lambda=2$) so that the result agrees with the canonical quantization result, which can be verified experimentally. Since $\alpha<1$, this means that, for our formalism to be in agreement with canonical quantization, we need to scale back the degree to which the fluctuation of the fermionic field causes the probability distribution to deviate from its original distribution.


\subsection{On the Schrödinger Picture}
The Schrödinger picture offers several advantages over the standard Fock space formulation of quantum fields~\cite{Long}. In particular, the Schrödinger wave functional provides an intrinsic description of the vacuum state without reference to the spectrum of excited states. This is especially significant in curved spacetime, where defining a unique vacuum in the Fock space formalism presents inherent challenges~\cite{Long}. Furthermore, the Schrödinger picture is often regarded as the most natural representation from the perspective of canonical quantum gravity, where spacetime is typically decomposed into a spatial manifold evolving in time~\cite{Corichi}. By formulating quantum field theory in the Schrödinger representation, we gain deeper insight into the similarities and differences between non-relativistic quantum mechanics and relativistic quantum field theory. This perspective may also offer new approaches for applying concepts from one framework to the other. For example, computing information-theoretic quantities such as entanglement entropy in quantum field theory remains a challenge~\cite{Takayanagi}. In non-relativistic quantum mechanics, entanglement entropy for a system is typically calculated using the wave function. With the availability of the Schrödinger wave functional in field theory, a similar methodology may be developed to compute the entanglement entropy in quantum field systems. 

\subsection{Limitations}
The assumption of field fluctuations serves as the foundation for defining the information metric $I_f$, which ultimately gives rise to the quantum behavior of the field. However, we do not provide a concrete physical model for these fluctuations. The underlying physics governing field fluctuations is expected to be complex and may hold the key to a deeper understanding of quantum field theory. Exploring this in detail is beyond the scope of this paper. Our goal here is to minimize the number of assumptions required to derive the Schrödinger equation for the wave functional, allowing future research to focus on justifying and refining these assumptions. 

The formulations presented in this paper are developed within a flat Minkowski spacetime. However, we expect that this framework can be extended to curved spacetime, enabling the derivation of the Schrödinger equation on a gravitational background. This remains an interesting direction for future exploration. 

It is also worth to point out that the results in Section \ref{sec:NLSE} are preliminary. It is not clear that quantizing a non-renormalizable fermionic field always leads to a non-linear Schr\"{o}dinger equation. More extensive investigation is needed to confirm the rigorousness and generality of the results.

\subsection{Conclusions}
In this paper, we have developed a quantization framework for fermionic fields based on the extended stationary action principle. Originally introduced to derive non-relativistic quantum theory~\cite{Yang2023} and later applied to scalar field quantization, this principle provides a novel perspective on the transition from classical to quantum field theory. By addressing the mathematical challenges of functional variation with Grassmann variables, we successfully derived the Floreanini-Jackiw representation of the Schrödinger equation for the wave functional. Furthermore, we verified that the resulting Hamiltonian operator generates the Poincaré algebra, ensuring that the theory maintains the full symmetry structure required by special relativity.

The extended stationary action principle offers a unique information-theoretic perspective on quantum field theory. As described in Section \ref{sec:LIP}, this framework is built on two fundamental assumptions. Assumption 2 establishes that the Planck constant defines the minimal discrete unit of action necessary for a field configuration to exhibit observable dynamics. In the classical limit, where this discrete action is effectively zero, the theory reduces to a classical field theory. Assumption 1 introduces a new metric, based on relative entropy, to quantify additional observable information arising from field fluctuations. This additional information metric is then converted to a correction term for the classical action via Assumption 2, leading to quantum behavior. By incorporating these entropy-based corrections into the Lagrangian, the classical field theory naturally transitions into a quantum field theory.

Our quantization framework serves as an alternative to the standard canonical quantization and path integral formulation. We verify that it reproduces the results of the canonical quantum field theory for fermions with a concrete calculation of the probability of particle creation under the influence of an external constant electric field. In addition, the framework offers an alternative approach for quantizing non-renormalizable theories, as demonstrated in Section \ref{sec:interactions}. Although renormalizable theories always admit a linear Schrödinger representation~\cite{Symanzik}, non-renormalizable theories do not necessarily possess such a representation. In particular, we showed that applying this framework to the non-renormalizable weak interaction between fermions leads to a nonlinear Schrödinger equation, a preliminary result that highlights the potential of this approach.

The works in Refs.~\cite{Yang2023, Yang2024, Yang2025}, along with the present study, demonstrate the flexibility and broad applicability of the mathematical framework based on the extended stationary action principle across both non-relativistic quantum mechanics and relativistic quantum field theory. Extending this framework to curved spacetime is highly feasible, providing a promising direction for future research. Since existing quantization methods face significant challenges in quantizing the gravitational field, exploring alternative approaches is desirable. Given the success of this framework in quantizing both scalar and fermionic fields, a natural next step is to investigate its applicability to quantizing the gravitational field, which is a topic for future study.

\begin{acknowledgements}
The author thanks the anonymous referees for their valuable comments, which lead to a better presentation of the paper, a stronger comparison of our results with the canonical quantization results in Section VII, and a clearer explanation of the physical meanings of the order of Tsallis divergence.
\end{acknowledgements}








\onecolumngrid

\pagebreak

\appendix

\section{Canonical Transformation for Fermionic Fields}
\label{appendix:canonical}
Suppose we choose a foliation of the Minkowski spacetime into a succession of fixed $t$ spacetime hypersurfaces $\Sigma_{t}$. The field configurations $(\psi, \psi^\dagger)$ for $\Sigma_{t}$ can be understood as a vector with infinitely many components for each spatial point on the Cauchy hypersurface $\Sigma_t$ at time instance $t$ and, denoted as $\psi_{t,\mathbf{x}}=\psi(t,\mathbf{x}) = \psi(x)$, and $\psi^\dagger_{t,\mathbf{x}}=\psi^\dagger(x)$. Here, the meaning of $\psi(x)$ should be understood as the field component $\psi_{\mathbf{x}}$ at each spatial point of the hypersurfaces $\Sigma_{t}$ at time instance $t$. We want to transform from the pairs of canonical variables $(\psi, \pi_\psi)$ and $(\psi^\dagger, \pi_{\psi^\dagger})$ into a generalized canonical variables $(\Phi, \Pi)$ and $(\Phi^\dagger, \Pi^\dagger)$, and preserve the form of canonical equations. Recall that we need to consider these fields variables as Grassmann-valued variables. Denote the Lagrangian for both canonical variables as $L=\int_{\Sigma_t} (\dot\psi\pi_\psi  + \dot\psi^\dagger\pi_{\psi^\dagger}) d^3x-H(\psi, \psi^\dagger, \pi_\psi,\pi_{\psi^\dagger})$ and $L'=\int_{\Sigma_t} (\dot\Phi\Pi+\dot\Phi^\dagger\Pi^\dagger) d^3x-K(\Phi, \Phi^\dagger,\Pi, \Pi^\dagger)$, respectively, where $H$ is defined in (\ref{H1}) and $K$ is the new form of Hamiltonian with the generalized canonical variables. We will omit the subscript ${\Sigma_t}$ in the integral. To ensure the form of canonical equations is preserved from the stationary action principle, one must have 
\begin{align}
    \delta \int dt L &= \delta\int dt \{\int(\dot\psi\pi_\psi  + \dot\psi^\dagger\pi_{\psi^\dagger}) d^3x-H(\psi, \psi^\dagger, \pi_\psi,\pi_{\psi^\dagger})\} = 0\\
    \delta \int dt L' &= \delta\int dt \{\int (\dot\Phi\Pi+\dot\Phi^\dagger\Pi^\dagger) d^3x-K(\Phi, \Phi^\dagger,\Pi, \Pi^\dagger)\} = 0.
\end{align}
One way to meet such conditions is that the Lagrangian in both integrals satisfy the following relation
\begin{equation}
    \label{extCan}
   \int (\dot\Phi\Pi+\dot\Phi^\dagger\Pi^\dagger) d^3x-K(\Phi, \Phi^\dagger,\Pi, \Pi^\dagger) = \lambda (\int(\dot\psi\pi_\psi  + \dot\psi^\dagger\pi_{\psi^\dagger}) d^3x-H(\psi, \psi^\dagger, \pi_\psi,\pi_{\psi^\dagger})) + \frac{dG}{dt},
\end{equation}
where $G$ is a generating functional, and $\lambda$ is a constant. When $\lambda \ne 1$, the transformation is called an extended canonical transformation. Re-arranging (\ref{extCan}), we have
\begin{equation}
    \label{extCan2}
    \frac{dG}{dt} = \int (\dot\Phi\Pi+\dot\Phi^\dagger\Pi^\dagger -\lambda (\dot\psi\pi_\psi  + \dot\psi^\dagger\pi_{\psi^\dagger})) d^3x - (K-\lambda H).
\end{equation}
Choose a generating functional $G=\int (\Phi\Pi+\Phi^\dagger\Pi^\dagger) d^3x - S(\psi, \psi^\dagger, \Pi, \Pi^\dagger, t)$, that is, a type 2 generating functional analogous to the type 2 generating function in classical mechanics~\cite{Yang2023}. Its total time derivative is
\begin{equation}
    \label{type2}
    \frac{dG}{dt} = \int (\dot\Phi\Pi+\dot\Phi^\dagger\Pi^\dagger + \Phi\dot\Pi +\Phi^\dagger\dot{\Pi}^\dagger) d^3x - \frac{\partial S}{\partial t} - \int (\dot{\psi}\frac{\delta S}{\delta\psi}+\dot{\psi}^\dagger\frac{\delta S}{\delta\psi^\dagger}+\dot{\Pi}\frac{\delta S}{\delta\Pi}+\dot{\Pi}^\dagger\frac{\delta S}{\delta\Pi^\dagger}) d^3x .
\end{equation}
Comparing (\ref{extCan2}) and (\ref{type2}) results in
\begin{align}
    \label{type12}
    \frac{\partial S}{\partial t} &=  K - \lambda H, \\
    \frac{\delta S}{\delta\psi} &=\lambda\pi_\psi, \text{   } \frac{\delta S}{\delta\psi^\dagger}=\lambda\pi_{\psi^\dagger},\\
    \frac{\delta S}{\delta\Pi} &= -\Phi,\text{  } \frac{\delta S}{\delta\Pi^\dagger} = -\Phi^\dagger.
\end{align}
From (\ref{type12}), $K=  (\partial S/\partial t +\lambda  H)$. Thus, $L'=\int (\dot\Phi\Pi+\dot\Phi^\dagger\Pi^\dagger)d^3x - (\partial S/\partial t +\lambda H)$. We can choose a generating functional $S$ such that $\Phi$ and $\Phi^\dagger$ do not explicitly depend on $t$ during motion so that $\dot{\Phi}=\dot{\Phi}^\dagger=0$ and $L' = -( \partial S/\partial t +\lambda H)$. Then the action functional with the generalized canonical variables becomes
\begin{equation}
    \label{extAction}
    A_c = \int dt L' = -\int dt \{\frac{\partial S}{\partial t} +\lambda H(\psi, \psi^\dagger,\pi, \pi^\dagger)\}.
\end{equation}
where the Hamiltonian $H$ is given in (\ref{H1}) or \eqref{H2}. If one further imposes constraint on the generating functional $S$ such that the generalized Hamiltonian $K=0$, Eq. (\ref{type12}) becomes the field theory version of the Hamilton-Jacobi equation for the functional $S$, $\partial S/\partial t + H = 0$ if we choose $\lambda=1$. 

Now consider that the field configurations $[\psi(x), \psi^\dagger(x)]$ are not definite but follow a probability distribution at any point of $\Sigma_t$. Alternatively, they can be understood as an ensemble of field configurations with probability density $\rho(\psi(x), \psi^\dagger(x), t)$. In this case, the Lagrangian density is $\rho L'$, and the total action functional for the ensemble of field configurations is,
\begin{equation}
    \label{extAction}
    S_c = -\int \mathcal{D}\psi^\dagger\mathcal{D}\psi dt \{\rho(\psi,\psi^\dagger, t)[\frac{\partial S}{\partial t} +\lambda H(\psi, \psi^\dagger,\pi, \pi^\dagger)]\},
\end{equation}
If we change the generalized canonical pair as $(\rho, S)$, applying the stationary action principle based on $S_c$ by variation of $S_c$ over $\rho$, one obtains, again, the field theory version of Hamilton-Jacobi equation for the functional $S$, $\partial S/\partial t + H = 0$.

\section{Inner Product and Expectation Value with Grassmann Variables}
\label{appendix:exp}
A fixed-time functional $\Psi[\omega, \omega^\dagger]$ can be viewed as a ket: $|\Psi\rangle \leftrightarrow \Psi[\omega, \omega^\dagger]$. The inner product is defined by functional integration
\begin{equation}
    \label{inner}
    \langle\Psi_1|\Psi_2\rangle = \int\mathcal{D}\omega^\dagger\mathcal{D}\omega\Psi_1^*\Psi_2.
\end{equation}
The dual functional $\langle\Psi| \leftrightarrow \Psi^*[\omega, \omega^\dagger]$, with Grassmann variables $\omega, \omega^\dagger$, is defined as~\cite{Kiefer, Jackiw2}
\begin{equation}
    \label{dual}
    \Psi^*[\omega, \omega^\dagger] = \int\mathcal{D}\bar{\omega}^\dagger\mathcal{D}\bar{\omega}\exp{(\bar{\omega}^\dagger\omega - \omega^\dagger\bar{\omega})}\bar{\Psi}[\bar{\omega},\bar{\omega}^\dagger],
\end{equation}
where $\bar{\Psi}$ is the Hermitian conjugate of $\Psi$. The same compact notation as \eqref{compact} is used for $\omega^\dagger\bar{\omega}\equiv\int dxdy \omega_\alpha^\dagger(y)\bar{\omega}_\alpha(x)$. The expectation value of $\omega_\alpha(x)\omega_\alpha^\dagger(y)$ is calculated as 
\begin{equation}
    \label{exp}
    \langle\omega_\alpha(x)\omega_\beta^\dagger(y)\rangle = \int\mathcal{D}\omega^\dagger\mathcal{D}\omega\Psi^*\omega_\alpha\omega_\beta^\dagger\Psi.
\end{equation}
Given the probability density in \eqref{transP2}, we define $\Psi = \sqrt{p}=\exp{(\Delta t\omega^\dagger h \omega)}$ (omitting the normalization factor $Z$). Denote $\Omega=h\Delta t$, it becomes
\begin{equation}
    \Psi = \exp{(\omega^\dagger \Omega \omega)}.
\end{equation}
The dual functional, by the rules of Grassmann integration, becomes~\cite{Kiefer}
\begin{equation}
    \Psi^* = \det(-\Omega^\dagger)\exp{(\omega^\dagger (\Omega^\dagger)^{-1} \omega)}.
\end{equation}
Note that the minus sign in $\det(-\Omega)$ arises because of the order of the integral measure $\mathcal{D}\omega^\dagger\mathcal{D}\omega$. The normalization factor becomes
\begin{equation}
\label{Normalization}
    \langle\Psi|\Psi\rangle = \det(-\Omega^\dagger)\int\mathcal{D}\omega^\dagger\mathcal{D}\omega\exp{(\omega^\dagger [\Omega+(\Omega^\dagger)^{-1}] \omega)} = \det (\Omega^\dagger\Omega+1).
\end{equation}
The normalized expectation value of $\omega_\alpha(x)\omega_\alpha^\dagger(y)$ is
\begin{equation}
    \label{exp}
    \begin{split}  
    \langle\omega_\alpha(x)\omega_\beta^\dagger(y)\rangle &= \frac{\det(-\Omega^\dagger)}{\det(\Omega^\dagger\Omega+1)}\int\mathcal{D}\omega^\dagger\mathcal{D}\omega\omega_\alpha\omega_\beta^\dagger\exp{(\omega^\dagger [\Omega+(\Omega^\dagger)^{-1}] \omega)}\\
    &=\frac{\det(-\Omega^\dagger)}{\det(\Omega^\dagger\Omega+1)}\frac{\delta^2}{\delta\eta_\beta\delta\eta^\dagger_\alpha}\int\mathcal{D}\omega^\dagger\mathcal{D}\omega\exp{(\omega^\dagger [\Omega+(\Omega^\dagger)^{-1}] \omega+\omega^\dagger\eta+\eta^\dagger\omega)|_{\eta=\eta^\dagger=0}}\\
    &=-(\Omega+(\Omega^\dagger)^{-1})^{-1}_{\alpha\beta}(x,y).
    \end{split}
\end{equation}
Substitute $\Omega=h\Delta t$ into the above equation, and note that $h$ is hermitian,
\begin{equation}
    \langle\omega_\alpha(x)\omega_\beta^\dagger(y)\rangle = -(\frac{h\Delta t}{(h\Delta t)^2 +1 })_{\alpha\beta}(x,y).
\end{equation}
When $\Delta t \to 0$, this is simplified as
\begin{equation}
    \langle\omega_\alpha(x)\omega_\beta^\dagger(y)\rangle = -h_{\alpha\beta}(x,y)\Delta t.
\end{equation}
For a general probability density $\rho[\omega,\omega^\dagger]$, the normalization factor $Z$ and expectation value for variable $O$ are
\begin{align}
    (\sqrt{\rho})^*&= \int\mathcal{D}\bar{\omega}^\dagger\mathcal{D}\bar{\omega}\exp{(\bar{\omega}^\dagger}\omega - \omega^\dagger\bar{\omega})\sqrt{\rho[\bar{\omega}, \bar{\omega}^\dagger]},\\
    \label{Normal}
    N &= \int\mathcal{D}\omega^\dagger\mathcal{D}\omega (\sqrt{\rho})^*(\sqrt{\rho}),\\
    \langle O\rangle& = \frac{1}{N}\int\mathcal{D}\omega^\dagger\mathcal{D}\omega (\sqrt{\rho})^*O(\sqrt{\rho}).
\end{align}

\section{Information Metrics for Field Fluctuations}
\label{appendix:I_f}
To derive (\ref{FisherInfo}) from (\ref{DLDivergence}) we need to take the functional derivative of $\rho[\psi+\omega, \psi^\dagger+\omega^\dagger, t_i]$ around $\psi$ and $\psi^\dagger$. But first we should be cautious about the correct formula for a Taylor expansion with Grassmann variable. For instance, let $f(u_1,u_2)=a+bu_1+cu_2+du_1u_2$ be a function with Grassmann variables $u_1$ and $u_2$. One can verify that the correct Taylor expansion is
\begin{equation}
    \label{Taylor}
    f(u_1+v_1, u_2+v_2) = f(u_1, u_1) + v_1\frac{\partial f}{\partial u_1}+ v_2\frac{\partial f}{\partial u_2} + v_1v_2\frac{\partial^2 f}{\partial u_1\partial u_2},
\end{equation}
instead of
\begin{equation}
    \label{Taylor2}
    f(u_1+v_1, u_2+v_2) = f(u_1, u_1) + \frac{\partial f}{\partial u_1}v_1+ \frac{\partial f}{\partial u_2}v_2 + \frac{\partial^2 f}{\partial u_1\partial u_2}v_1v_2.
\end{equation}
With this in mind, let us expand $\rho[\psi+\omega, \psi^\dagger+\omega^\dagger, t_i]$ up to the second order. We will omit the time labeling for $\rho$.
\begin{equation}
\label{Taylor}
    \rho[\psi+\omega, \psi^\dagger+\omega^\dagger] = \rho[\psi,\psi^\dagger] + \int d^3x\omega_\alpha(x)\frac{\delta\rho}{\delta\psi_\alpha(x)} +\int d^3y\omega_\beta^\dagger(y)\frac{\delta\rho}{\delta\psi_\beta^\dagger(y)}+\int d^3xd^3y\omega_\alpha(x)\omega^\dagger_\beta(y)\frac{\delta^2\rho}{\delta\psi_\beta^\dagger(y)\delta\psi_\alpha(x)} .
\end{equation}
Note the convention of Einstein summation on the field component indices $\alpha, \beta$.
The expansion is legitimate because (\ref{expectation}) shows that the expectation value of fluctuation displacement $\omega_\alpha\omega^\dagger_\beta$ is proportional to $\Delta t$. As $\Delta t \to 0$, only very small $\omega$ and $\omega^\dagger$ are significant. Then
\begin{align*}
\label{Taylor2}
    ln\frac{\rho[\psi+\omega, \psi^\dagger+\omega^\dagger]}{\rho[\psi,\psi^\dagger]} = & ln \{1 + \frac{1}{\rho}[\int d^3x\omega_\alpha(x)\frac{\delta\rho}{\delta\psi_\alpha(x)} + \int d^3y\frac{\delta\rho}{\delta\psi_\beta^\dagger(y)}\omega_\beta^\dagger(y)+\int d^3xd^3y\omega_\alpha(x)\omega^\dagger_\beta(y)\frac{\delta^2\rho}{\delta\psi_\beta^\dagger(y)\delta\psi_\alpha(x)} ] \\
     =& \frac{1}{\rho}[\int d^3x\omega_\alpha(x)\frac{\delta\rho}{\delta\psi_\alpha(x)} + \int d^3y\omega_\beta^\dagger(y)\frac{\delta\rho}{\delta\psi_\beta^\dagger(y)}+\int d^3xd^3y\omega_\alpha(x)\omega^\dagger_\beta(y)\frac{\delta^2\rho}{\delta\psi_\beta^\dagger(y)\delta\psi_\alpha(x)} ] \\
    &-\frac{1}{2\rho^2}[\int d^3x\omega_\alpha(x)\frac{\delta\rho}{\delta\psi_\alpha(x)} + \int d^3y\omega_\beta^\dagger(y)\frac{\delta\rho}{\delta\psi_\beta^\dagger(y)}]^2
\end{align*}
Substitute the above expansion into (\ref{DLDivergence}), and take the expectation values $\langle\cdot\rangle_{\omega}$. Owning to the identities in \eqref{expectation} and \eqref{expectation2}, the only surviving terms are
\begin{align*}
    &\langle D_{KL}(\rho[\psi, \psi^\dagger, t_i] || \rho[\psi+\omega,\psi^\dagger+\omega^\dagger, t_i])\rangle_\omega \\
    &= -\int\mathcal{D}\psi^\dagger\mathcal{D}\psi\{\int d^3xd^3y\langle\omega_\alpha(x)\omega^\dagger_\beta(y)\rangle\frac{\delta^2\rho}{\delta\delta\psi_\beta^\dagger(y)\psi_\alpha(x)}   +\int d^3xd^3y\frac{1}{\rho}\frac{\delta\rho}{\delta\psi_\alpha(x)}\langle\omega_\alpha(x)\omega_\beta^\dagger(y)\rangle_\omega\frac{\delta\rho}{\delta\psi_\beta^\dagger(y)} \}\\
    &= \Delta t\int\mathcal{D}\psi^\dagger\mathcal{D}\psi\int d^3xd^3y\{h_{\alpha\beta}(x,y)\frac{\delta^2\rho}{\delta\psi_\alpha(x)\delta\psi_\beta^\dagger(y)}+\frac{1}{\rho}\frac{\delta\rho}{\delta\psi_\alpha(x)}h_{\alpha\beta}(x,y)\frac{\delta\rho}{\delta\psi_\beta^\dagger(y)}\}
\end{align*}
Performing the integration in the first term by explicitly expanding the integration measure $\mathcal{D}\psi^\dagger\mathcal{D}\psi$ over all the spatial points $x,y$ in the hypersurface $\Sigma_{t_i}$,
\begin{align}
    \int d^3xd^3y\mathcal{D}\psi^\dagger\mathcal{D}\psi\frac{\delta^2\rho}{\delta\psi(x)\delta\psi^\dagger(y)} &= \sum_{x,y\in\Sigma_{t_i}}\int \prod_{x',y'\in\Sigma_{t_i}} d\psi_{x'}^\dagger d\psi_{x'} \frac{\delta}{\delta\psi_{x}}(\frac{\delta\rho}{\delta \psi_{y}^\dagger}) \\
    \label{smoothRho}
    &=\sum_{x,y\in\Sigma_{t_i}}\int \prod_{x'\ne x, y'\ne y} d\psi_{x'}^\dagger d\psi_{x'}(\rho\vert_{\psi_{x},\psi^\dagger_y=\infty} - \rho\vert_{\psi_{x},\psi^\dagger_y=-\infty}).
\end{align}
We temporarily omit the component label $\alpha,\beta$ in the above integral. Assuming $\rho$ is a smooth functional such that it approaches zero when $\psi_{x},\psi^\dagger_y$ approaches the boundary, the above integral vanishes. Thus,
\begin{equation}
    \langle D_{KL}(\rho[\psi, \psi^\dagger, t_i] || \rho[\psi+\omega,\psi^\dagger+\omega^\dagger, t_i])\rangle_{\omega} = \Delta t\int\mathcal{D}\psi^\dagger\mathcal{D}\psi\int d^3xd^3y\frac{1}{\rho}\frac{\delta\rho}{\delta\psi_\alpha(x)}h_{\alpha\beta}(x,y)\frac{\delta\rho}{\delta\psi_\beta^\dagger(y)}.
\end{equation}
Substitute this into (\ref{DLDivergence}),
\begin{align}
    \label{totalInfo4}
    I_f &= \sum_{i=0}^{N-1}\langle D_{KL}(\rho[\psi, \psi^\dagger, t_i] || \rho[\psi+\omega,\psi^\dagger+\omega^\dagger, t_i])\rangle_{\omega} = \int dt\int\mathcal{D}\psi^\dagger\mathcal{D}\psi\int d^3xd^3y\frac{1}{\rho}\frac{\delta\rho}{\delta\psi_\alpha(x)}h_{\alpha\beta}(x,y)\frac{\delta\rho}{\delta\psi_\beta^\dagger(y)}.
\end{align}
Written in matrix format, it becomes (\ref{FisherInfo}). 

\section{Integration by Parts with Grassmann Variables}
\label{Appendix:IntByPart}
Denote $f(u, u^\dagger)$ and $g(u,u^\dagger)$ are two functions with Grassmann variable $u$ and $u^\dagger$. Swapping the order between $u$ and $f(u, u^\dagger)$ produces the following result:
\begin{equation}
    uf(u, u^\dagger) \rightarrow f(-u, -u^\dagger)u.
\end{equation}
Similarly, since the derivative $\frac{\partial}{\partial u}$ itself is considered a Grassmann variable, swapping the order between $\frac{\partial}{\partial u}$ and $f(u, u^\dagger)$ gives
\begin{equation}
    \frac{\partial}{\partial u}f(u, u^\dagger) \rightarrow f(-u, -u^\dagger)\frac{\partial}{\partial u}.
\end{equation}
Since
\begin{equation}
    0=\int du^\dagger du \frac{\partial}{\partial u}(fg) = \int du^\dagger du (\frac{\partial}{\partial u}f)g + \int du^\dagger du f(-u,-u^\dagger)\frac{\partial}{\partial u}(g).
\end{equation}
We have
\begin{align}
    \int du^\dagger du (\frac{\partial}{\partial u}f(u,u^\dagger))g(u,u^\dagger) &=- \int du^\dagger du f(-u,-u^\dagger)\frac{\partial}{\partial u}g(u,u^\dagger).
\end{align}
Applying the same logic to functional $F(\phi,\phi^\dagger)$ and $G(\phi,\phi^\dagger)$, where $\phi(x)$ and $\phi^\dagger(y)$ are Grassmann-valued fields
\begin{align}
    \int \mathcal{D}\phi^\dagger\mathcal{D}\phi (\frac{\delta}{\delta \phi}F(\phi,\phi^\dagger))G(\phi,\phi^\dagger) =- \int \mathcal{D}\phi^\dagger\mathcal{D}\phi F(-\phi,-\phi^\dagger)\frac{\delta}{\delta \phi}G(\phi,\phi^\dagger).
\end{align}
Next we generalize to multi-component Grassmann-valued fields $\psi$, $\psi^\dagger$ with components $\psi_\alpha$ and $\psi_\beta^\dagger$
\begin{equation}
\label{D6}
    \int \mathcal{D}\psi^\dagger\mathcal{D}\psi (\frac{\delta}{\delta \psi_\alpha}F(\psi,\psi^\dagger))\Omega_{\alpha\beta}G_\beta(\psi,\psi^\dagger) =- \int \mathcal{D}\psi^\dagger\mathcal{D}\psi F(-\psi,-\psi^\dagger)\frac{\delta}{\delta \psi_\alpha}\Omega_{\alpha\beta}G_\beta(\psi,\psi^\dagger).
\end{equation}
where $\Omega_{\alpha\beta}$ is an element of the matrix $\Omega$. Let $F=\delta'S(\psi,\psi^\dagger)$ where $\delta'$ represents a small variation of functional $S$, and $G=\frac{\delta T}{\delta\psi_\beta^\dagger}$, the above equation becomes
\begin{equation}
    \int \mathcal{D}\psi^\dagger\mathcal{D}\psi (\frac{\delta}{\delta \psi_\alpha}\delta'S(\psi,\psi^\dagger))\Omega_{\alpha\beta}\frac{\delta T(\psi,\psi^\dagger)}{\delta\psi^\dagger_\beta} =- \int \mathcal{D}\psi^\dagger\mathcal{D}\psi (\delta'S(-\psi,-\psi^\dagger))\frac{\delta}{\delta \psi_\alpha}\Omega_{\alpha\beta}\frac{\delta T(\psi,\psi^\dagger)}{\delta\psi^\dagger_\beta}.
\end{equation}
Let $\Omega_{\alpha\beta}=h_{\alpha\beta}$, and rewrite the equation above in a more compact matrix form,
\begin{equation}
\label{IBP-1}
    \int \mathcal{D}\psi^\dagger\mathcal{D}\psi (\frac{\delta}{\delta \psi}\delta'S(\psi,\psi^\dagger))h\frac{\delta T(\psi,\psi^\dagger)}{\delta\psi^\dagger} =- \int \mathcal{D}\psi^\dagger\mathcal{D}\psi (\delta'S(-\psi,-\psi^\dagger))\frac{\delta}{\delta \psi}h\frac{\delta T(\psi,\psi^\dagger)}{\delta\psi^\dagger}.
\end{equation}   
Similarly, in \eqref{D6}, if we let $G=\delta'S$ and $F=\frac{\delta T}{\delta \psi_\beta^\dagger}$, we obtain
\begin{align}
    \int \mathcal{D}\psi^\dagger\mathcal{D}\psi \frac{\delta T(\psi,\psi^\dagger)}{\delta \psi_\alpha}\Omega_{\alpha\beta}\frac{\delta}{\delta\psi^\dagger_\beta}(\delta'S(\psi,\psi^\dagger)) &=- \int \mathcal{D}\psi^\dagger\mathcal{D}\psi \frac{\delta}{\delta\psi^\dagger_\beta}\Omega_{\alpha\beta}\frac{\delta T(-\psi,-\psi^\dagger)}{\delta(-\psi_\alpha)}(\delta'S(\psi,\psi^\dagger))\\
    &=- \int \mathcal{D}\psi^\dagger\mathcal{D}\psi \frac{\delta}{\delta\psi_\alpha}\Omega_{\alpha\beta}\frac{\delta T(-\psi,-\psi^\dagger)}{\delta\psi^\dagger_\beta}(\delta'S(\psi,\psi^\dagger)).
\end{align}
In matrix format, this is
\begin{equation}
\label{IBP-2}
    \int \mathcal{D}\psi^\dagger\mathcal{D}\psi \frac{\delta T(\psi,\psi^\dagger)}{\delta \psi}h\frac{\delta}{\delta\psi^\dagger}(\delta'S(\psi,\psi^\dagger)) = - \int \mathcal{D}\psi^\dagger\mathcal{D}\psi \frac{\delta}{\delta\psi}h\frac{\delta T(-\psi,-\psi^\dagger)}{\delta\psi^\dagger}(\delta'S(\psi,\psi^\dagger)).
\end{equation}
If the functional $T$ is invariant with changing signs of the field variables, that is, $T(-\psi,-\psi^\dagger) = T(\psi,\psi^\dagger)$, the rules for integration by part, \eqref{IBP-1} and \eqref{IBP-2}, are the same as those with regular non-Grassmann variables. Fortunately, the Lagrangian for fermionic fields is always coupling $\psi$ with $\psi^\dagger$, that is, $\psi$ always appears in pair with $\psi^\dagger$ for each term in the Lagrangian. We expect that the functionals $S$ and $\rho$ can consist of all possible combinations in terms of pairs $(\psi_\alpha^\dagger\psi_\beta)$. However, flipping the signs for both $\psi_\beta$ and $\psi^\dagger_\alpha$ at the same time does not result in a change in sign. Thus, we can safely assume $S(-\psi,-\psi^\dagger) = S(\psi,\psi^\dagger)$, and $\rho(-\psi,-\psi^\dagger) = \rho(\psi,\psi^\dagger)$ in the rest of this paper. This greatly simplifies the integration by part in our calculations.

\section{Derivation of the Schr\"{o}dinger Equation}
\label{appendix:SE}
To derive equation \eqref{ContEq}, we perform the variation procedure on \eqref{totalDist} with respect to $S$. The first term becomes
\begin{equation}
   - \delta'\int dt\mathcal{D}\psi^\dagger\mathcal{D}\psi(\rho\frac{\partial S}{\partial t} ) =  \int dt\mathcal{D}\psi^\dagger\mathcal{D}\psi(\frac{\partial \rho}{\partial t} \delta'S).
\end{equation}
For the variation of the second term, we need to use \eqref{IBP-1} and \eqref{IBP-2},
\begin{align}
    &\delta' \int dt\mathcal{D}\psi^\dagger\mathcal{D}\psi\int d^3xd^3y (\frac{4\rho}{\lambda}\frac{\delta S}{\delta\psi}h\frac{\delta S}{\delta\psi^\dagger}) = \frac{4}{\lambda}\int dt\mathcal{D}\psi^\dagger\mathcal{D}\psi\int d^3xd^3y \{\rho\frac{\delta (\delta'S)}{\delta\psi}h\frac{\delta S}{\delta\psi^\dagger}+\rho\frac{\delta S}{\delta\psi}h\frac{\delta (\delta'S)}{\delta\psi^\dagger}\} \\
    &=-\frac{4}{\lambda}\int dt\mathcal{D}\psi^\dagger\mathcal{D}\psi\int d^3xd^3y \{(\frac{\delta \rho}{\delta\psi}h\frac{\delta S}{\delta\psi^\dagger}+\rho\frac{\delta }{\delta\psi}h\frac{\delta S}{\delta\psi^\dagger}) + (\frac{\delta S}{\delta\psi}h\frac{\delta \rho}{\delta\psi^\dagger}+\rho\frac{\delta }{\delta\psi}h\frac{\delta S}{\delta\psi^\dagger})\}\delta'S.
\end{align}
Note that the symbol $\delta'$ refers to the variation over the functional $S$ while $\delta$ refers to the variation over the field variable $\psi$. The third term in \eqref{totalDist} vanishes when we take variation with respect to $S$. Combining the above two results, and demanding $\delta'S_t = 0$ for arbitrary $\delta'S$, we obtain
\begin{align}
    \frac{\partial\rho}{\partial t} &- \frac{4}{\lambda}\int d^3xd^3y \{\frac{\delta \rho}{\delta\psi}h\frac{\delta S}{\delta\psi^\dagger} + \frac{\delta S}{\delta\psi}h\frac{\delta \rho}{\delta\psi^\dagger} + 2\rho\frac{\delta }{\delta\psi}h\frac{\delta S}{\delta\psi^\dagger}\}=0.
\end{align}

The next step is to derive (\ref{varI_f}). Variation of $I_f$ given in (\ref{FisherInfo}) with a small arbitrary change of  $\rho$, $\delta'\rho$, results in
\begin{align}
    \delta' I_f =& \int dt\mathcal{D}\psi^\dagger\mathcal{D}\psi\int d^3xd^3y\{-\frac{\delta'\rho}{\rho^2}\frac{\delta \rho}{\delta\psi}h\frac{\delta \rho}{\delta\psi^\dagger}+\frac{1}{\rho}\frac{\delta (\delta'\rho)}{\delta\psi}h\frac{\delta \rho}{\delta\psi^\dagger} + \frac{1}{\rho}\frac{\delta \rho}{\delta\psi}h\frac{\delta (\delta'\rho)}{\delta\psi^\dagger}\}\\
    &=\int dt\mathcal{D}\psi^\dagger\mathcal{D}\psi\int d^3xd^3y\{-\frac{1}{\rho^2}\frac{\delta \rho}{\delta\psi}h\frac{\delta \rho}{\delta\psi^\dagger}-\frac{\delta }{\delta\psi}(\frac{1}{\rho}h\frac{\delta \rho}{\delta\psi^\dagger}) - \frac{\delta }{\delta\psi^\dagger}(\frac{1}{\rho}h^T\frac{\delta \rho}{\delta\psi})\}\delta'\rho\\
    \label{E7}
    &=\int dt\mathcal{D}\psi^\dagger\mathcal{D}\psi\int d^3xd^3y\{\frac{1}{\rho^2}\frac{\delta \rho}{\delta\psi}h\frac{\delta \rho}{\delta\psi^\dagger}-\frac{2}{\rho}\frac{\delta }{\delta\psi}h\frac{\delta \rho}{\delta\psi^\dagger}\}\delta'\rho.
\end{align}
Defining $R=\sqrt{\rho}$, one can verify that
\begin{equation}
    -\frac{4}{R}\frac{\delta }{\delta\psi}h\frac{\delta R}{\delta\psi^\dagger}=\frac{1}{\rho^2}\frac{\delta \rho}{\delta\psi}h\frac{\delta \rho}{\delta\psi^\dagger}-\frac{2}{\rho}\frac{\delta }{\delta\psi}h\frac{\delta \rho}{\delta\psi^\dagger}. 
\end{equation}
Inserting it into \eqref{E7} gives \eqref{varI_f}.

Now defining $\Psi[\phi, t]=\sqrt{\rho[\phi, t]}e^{iS}$, and substituting \eqref{QHJ} and the continuity equation (\ref{ContEq}), we have
\begin{align}
    \frac{i}{\Psi}\frac{\partial \Psi}{\partial t} =& \frac{i}{2\rho}\frac{\partial \rho}{\partial t} - \frac{\partial S}{\partial t}\\
    \label{SE11}
    =& \int d^3xd^3y \{\frac{4i}{\lambda}(\frac{1}{2\rho}\frac{\delta \rho}{\delta\psi}h\frac{\delta S}{\delta\psi^\dagger} + \frac{1}{2\rho}\frac{\delta S}{\delta\psi}h\frac{\delta \rho}{\delta\psi^\dagger} + \frac{\delta }{\delta\psi}h\frac{\delta S}{\delta\psi^\dagger})-(\frac{4}{\lambda}\frac{\delta S}{\delta\psi}h\frac{\delta S}{\delta\psi^\dagger}+\frac{1}{2\rho^2}\frac{\delta \rho}{\delta\psi}h\frac{\delta \rho}{\delta\psi^\dagger}-\frac{1}{\rho}\frac{\delta }{\delta\psi}h\frac{\delta \rho}{\delta\psi^\dagger})\}
\end{align}
On the other hand, computing the second order of functional derivative of $\Psi$ gives
\begin{align}
    \frac{\delta\Psi}{\delta\psi^\dagger} &= \frac{1}{2\rho}\frac{\delta\rho}{\delta\psi^\dagger}\Psi + i\frac{\delta S}{\delta\psi^\dagger}\Psi \\
    \frac{\delta }{\delta\psi}h\frac{\delta}{\delta\psi^\dagger} \Psi&= \{i(\frac{1}{2\rho}\frac{\delta \rho}{\delta\psi}h\frac{\delta S}{\delta\psi^\dagger} + \frac{1}{2\rho}\frac{\delta S}{\delta\psi}h\frac{\delta \rho}{\delta\psi^\dagger} + \frac{\delta }{\delta\psi}h\frac{\delta S}{\delta\psi^\dagger})-(\frac{\delta S}{\delta\psi}h\frac{\delta S}{\delta\psi^\dagger}+\frac{1}{4\rho^2}\frac{\delta \rho}{\delta\psi}h\frac{\delta \rho}{\delta\psi^\dagger}-\frac{1}{2\rho}\frac{\delta }{\delta\psi}h\frac{\delta \rho}{\delta\psi^\dagger})\}\Psi\\
    \label{SE12}
    \frac{4}{\lambda}\frac{\delta }{\delta\psi}h\frac{\delta}{\delta\psi^\dagger} \Psi &= \{\frac{4i}{\lambda}(\frac{1}{2\rho}\frac{\delta \rho}{\delta\psi}h\frac{\delta S}{\delta\psi^\dagger} + \frac{1}{2\rho}\frac{\delta S}{\delta\psi}h\frac{\delta \rho}{\delta\psi^\dagger} + \frac{\delta }{\delta\psi}h\frac{\delta S}{\delta\psi^\dagger})-(\frac{4}{\lambda}\frac{\delta S}{\delta\psi}h\frac{\delta S}{\delta\psi^\dagger}+\frac{1}{\lambda\rho^2}\frac{\delta \rho}{\delta\psi}h\frac{\delta \rho}{\delta\psi^\dagger}-\frac{2}{\lambda\rho}\frac{\delta }{\delta\psi}h\frac{\delta \rho}{\delta\psi^\dagger})\}\Psi.
\end{align}
Comparing (\ref{SE11}) and (\ref{SE12}), and choosing $\lambda=2$, we obtain the Schr\"{o}dinger equation for the wave functional $\Psi$,
\begin{equation}
    i\frac{\partial \Psi}{\partial t} = 2\int d^3xd^3y (\frac{\delta }{\delta\psi}h\frac{\delta}{\delta\psi^\dagger} )\Psi.
\end{equation}

\section{Tsallis Divergence}
\label{appendix:RE}
Based on the definition of $I_f^{\alpha}$ in (\ref{TDivergence}), and starting from (\ref{Taylor}), we have
\begin{align*}
    \int \mathcal{D}\psi^\dagger\mathcal{D}\psi \frac{\rho^\alpha[\psi, \psi^\dagger, t_i]}{\rho^{\alpha-1} [\psi+\omega, \psi^\dagger+\omega^\dagger,t_i]} =& \int \mathcal{D}\psi^\dagger\mathcal{D}\psi\rho(1+\frac{1}{\rho}[\int d^3x\omega_\alpha(x)\frac{\delta\rho}{\delta\psi_\alpha(x)} + \int d^3y\omega_\beta^\dagger(y)\frac{\delta\rho}{\delta\psi_\beta^\dagger(y)}])^{1-\alpha} \\
    =& \int\mathcal{D}\psi^\dagger\mathcal{D}\psi \{\rho+(1-\alpha)[\int d^3x\omega_\alpha(x)\frac{\delta\rho}{\delta\psi_\alpha(x)} + \int d^3y\omega_\beta^\dagger(y)\frac{\delta\rho}{\delta\psi_\beta^\dagger(y)}] \\
    & +\frac{1}{2}\alpha(\alpha-1)(\frac{1}{\rho}[\int d^3x\omega_\alpha(x)\frac{\delta\rho}{\delta\psi_\alpha(x)} + \int d^3y\omega_\beta^\dagger(y)\frac{\delta\rho}{\delta\psi_\beta^\dagger(y)}]^2) \}.
\end{align*}
Substitute the above expansion into (\ref{TDivergence}), and take the expectation values $\langle\cdot\rangle_{\omega}$. Due to the identities in \eqref{expectation} and \eqref{expectation2}, $I_f^{\alpha}$ is simplified as
\begin{align}
\label{RDivergence2}
    I_f^{\alpha} 
    &==\sum_{i=0}^{N-1}\langle\frac{1}{\alpha-1}(\int \mathcal{D}\psi^\dagger\mathcal{D}\psi \frac{\rho^\alpha[\psi, \psi^\dagger, t_i]}{\rho^{\alpha-1} [\psi+\omega, \psi^\dagger+\omega^\dagger,t_i]} - Z)\rangle_\omega. \\
    & = - \sum_{i=0}^{N-1}\alpha \int \mathcal{D}\psi^\dagger\mathcal{D}\psi\int d^3xd^3y\frac{1}{\rho}\frac{\delta\rho}{\delta\psi_\alpha(x)}\langle\omega_\alpha(x)\omega_\beta^\dagger(y)\rangle_\omega\frac{\delta\rho}{\delta\psi_\beta^\dagger(y)} \\
\label{I_f4}
    &= \alpha\int dt\mathcal{D}\psi^\dagger\mathcal{D}\psi\int d^3xd^3y\frac{1}{\rho}\frac{\delta\rho}{\delta\psi_\alpha(x)}h_{\alpha\beta}(x,y)\frac{\delta\rho}{\delta\psi_\beta^\dagger(y)} = \alpha I_f.
\end{align}

\section{Proof of the Poincar\'{e} Algebra}
\label{appendix:Poincare}
In this appendix, we will frequently encounter the following integral with derivative of the Dirac delta function
\begin{equation}
    I =\int\int dx dy f(x)g(y)\partial_y\delta(x-y).
\end{equation}
We can first proceed with integration of $y$, and perform integration by part,
\begin{equation}
\begin{split}
\label{IntDeltaDer}
    I &= \int dx f(x)[\int dy g(y)\partial_y\delta(x-y)] =\int dx f(x)[-\int dy \delta(x-y)\partial_yg(y)] \\
    & =-\int dx f(x)\partial_xg(x) = \int dx (\partial_xf(x)) g(x).
\end{split}
\end{equation}
For simplified notations, we write the Hamiltonian operator \eqref{JackiwHDO} as a linear combination of four terms, integrate the $\delta(x-y)$ function inside the operator $h$, and suppress the superscripts in $d^3xd^3y$,
\begin{subequations}
\begin{align}
    \hat{H} &= \frac{\lambda}{4}\hat{H}_1 + \frac{1}{2}\hat{H}_2+\frac{1}{2}\hat{H}_3+\frac{1}{\lambda}\hat{H}_4\\
    \hat{H}_1 &= \int dx \hat{\mathcal{H}}_1, \mbox{ }\hat{\mathcal{H}}_1 =u^\dagger h u\\
    \hat{H}_2 &= \int dx \hat{\mathcal{H}}_2, \mbox{ }\hat{\mathcal{H}}_2=(\frac{\delta}{\delta u}) h u\\
    \hat{H}_2 &= \int dx \hat{\mathcal{H}}_3, \mbox{ }\hat{\mathcal{H}}_3=u^\dagger h\frac{\delta}{\delta u^\dagger} \\
    \hat{H}_2 &= \int dx \hat{\mathcal{H}}_4, \mbox{ }\hat{\mathcal{H}}_4=\frac{\delta}{\delta u} h\frac{\delta}{\delta u^\dagger}.
\end{align}
\end{subequations}
Given the definition of $\hat{P}_i$ in \eqref{momentumOp}, we have
\begin{equation}
\label{Pcommutor}
    [\hat{P}_i, \hat{P}_j] =-i\int dxdy[(\partial_{ix}u^\dagger_x\frac{\delta}{\delta u_x^\dagger }+\partial_{ix}u_x\frac{\delta}{\delta u_x}), (\partial_{jy}u^\dagger_y\frac{\delta}{\delta u_y^\dagger }+\partial_{jy}u_y\frac{\delta}{\delta u_y})].
\end{equation}
Using \eqref{IntDeltaDer}, one obtains
\begin{align*}
    \int dxdy [\partial_{ix}u_x\frac{\delta}{\delta u_x}, \partial_{jy}u_y\frac{\delta}{\delta u_y}]&=\int dxdy\{\partial_{ix}u_x(\frac{\delta}{\delta u_x}\partial_{jy}u_y)\frac{\delta}{\delta u_y} - \partial_{jy}u_y(\frac{\delta}{\delta u_y}\partial_{ix}u_x)\frac{\delta}{\delta u_x}\}\\
    &=\int dxdy\{\partial_{ix}u_x(\partial_{jy}\delta(x-y))\frac{\delta}{\delta u_y} - \partial_{jy}u_y(\partial_{ix}\delta(x-y))\frac{\delta}{\delta u_x}\}\\
    &=\int dx\{(\partial_{jx}\partial_{ix}u_x)\frac{\delta}{\delta u_x} - (\partial_{ix}\partial_{jx}u_x)\frac{\delta}{\delta u_x}\}=0.
\end{align*}
Similarly,
\begin{equation*}
    \int dxdy [\partial_{ix}u_x^\dagger\frac{\delta}{\delta u_x^\dagger}, \partial_{jy}u_y^\dagger\frac{\delta}{\delta u_y^\dagger}]= 0.
\end{equation*}
On the other hand, since $\delta u^\dagger/\delta u = \delta u/\delta u^\dagger=0$,
\begin{equation*}
        \int dxdy [\partial_{ix}u_x^\dagger\frac{\delta}{\delta u_x^\dagger}, \partial_{jy}u_y\frac{\delta}{\delta u_y}]=\int dxdy [\partial_{ix}u_x\frac{\delta}{\delta u_x}, \partial_{jy}u_y^\dagger\frac{\delta}{\delta u_y^\dagger}]=0.
\end{equation*}    
Inserting the above three equations into \eqref{Pcommutor}, one gets $[\hat{P}_i, \hat{P}_j]=0$. 

To calculate $[\hat{P}_i, \hat{H}]$, we evaluate each of the four terms for $\hat{H}$,
\begin{align*}
    [\hat{P}_i, \hat{H}_1]&=-i\int dxdy (\partial_{ix}u^\dagger_x\frac{\delta}{\delta u_x^\dagger }+\partial_{ix}u_x\frac{\delta}{\delta u_x})u^\dagger_yhu_y =-i\int dx(\partial_{ix}u^\dagger_xhu_x + u_x^\dagger h\partial_{ix}u_x) =0,\\
    [\hat{P}_i, \hat{H}_2]&=-i\int dxdy \{(\partial_{ix}u_x)\frac{\delta}{\delta u_x}(\frac{\delta}{\delta u_y}) h u_y - (\frac{\delta}{\delta u_y}) h u_y(\partial_{ix}u_x)\frac{\delta}{\delta u_x}\} \\
    & =-i\int dxdy\{(\partial_{ix}u_x)(-\frac{\delta}{\delta u_y} h\delta(x-y) + h u_y(\partial_{ix}\delta(x-y)\frac{\delta}{\delta u_x}\} \\
    &=-i\int dx\{\frac{\delta}{\delta u_x} h(\partial_{ix}u_x) + h(\partial_{ix}u_x)\frac{\delta}{\delta u_x}\} = 0,\\
    [\hat{P}_i, \hat{H}_4] &=-i\int dxdy [(\partial_{ix}u^\dagger_x\frac{\delta}{\delta u_x^\dagger }+\partial_{ix}u_x\frac{\delta}{\delta u_x}), \frac{\delta}{\delta u_y} h\frac{\delta}{\delta u^\dagger_y}] =i\int dxdy \{\frac{\delta}{\delta u_y} h(\frac{\delta}{\delta u^\dagger_y}\partial_{ix}u^\dagger_x)\frac{\delta}{\delta u_x^\dagger}+(\frac{\delta}{\delta u_y} h\frac{\delta}{\delta u^\dagger_y}\partial_{ix}u_x)\frac{\delta}{\delta u_x}\}\\
    &=i\int dxdy\{\frac{\delta}{\delta u_y} h(\partial_{ix}\delta(x-y))\frac{\delta}{\delta u_x^\dagger}-h\frac{\delta}{\delta u^\dagger_y}(\partial_{ix}\delta(x-y))\frac{\delta}{\delta u_x}\}\\
    &=i\int dx\{(\partial_{ix}\frac{\delta}{\delta u_x} )h\frac{\delta}{\delta u_x^\dagger}-(\partial_{ix}h\frac{\delta}{\delta u^\dagger_x})\frac{\delta}{\delta u_x} \} = i\int dx\{(\partial_{ix}\frac{\delta}{\delta u_x} )h\frac{\delta}{\delta u_x^\dagger}+h\frac{\delta}{\delta u^\dagger_x}(\partial_{ix}\frac{\delta}{\delta u_x})\} =0.
\end{align*}
Note that the last step for $[\hat{P}_i, \hat{H}_1]=0$ uses the integration by part, and the last step of $[\hat{P}_i, \hat{H}_2]=0$ uses the properties of Grassmann variables. The proof of $[\hat{P}_i, \hat{H}_3]=0$ is not shown above, as it is similar to $[\hat{P}_i, \hat{H}_2]=0$. The linear combination of these commutators also holds. Thus, $[\hat{P}_i, \hat{H}]=0$. 

Similarly, to evaluate the commutator with the Lorentz boost operator, one can evaluate the following commutators,
\begin{align*}
    &[\hat{K}_i, \hat{P}_j] = \int dx [x_i\hat{\mathcal{H}}, \hat{P}_j] - t[\hat{P}_i, \hat{P}_j] = \frac{\lambda}{4}\int dx [x_i\hat{\mathcal{H}_1}, \hat{P}_j]+\frac{1}{2}\int dx [x_i\hat{\mathcal{H}}_2, \hat{P}_j]+\frac{1}{2}\int dx [x_i\hat{\mathcal{H}}_3, \hat{P}_j]+\frac{1}{\lambda}\int dx [x_i\hat{\mathcal{H}}_4, \hat{P}_j],\\
    &\int dx [x_i\hat{\mathcal{H}_1}, \hat{P}_j]=i\int dxdy \{(\partial_{jx}u^\dagger_x\frac{\delta}{\delta u_x^\dagger }+\partial_{jx}u_x\frac{\delta}{\delta u_x})(x_iu^\dagger_yhu_y)=i\int dx\{(\partial_{jx}u^\dagger_x)x_ihu_x + x_iu^\dagger_xh(\partial_{jx}u_x)\}=-i\delta_{ij}\hat{H}_1,\\
    &\int dx [x_i\hat{\mathcal{H}_2}, \hat{P}_j]=i\int dxdy \{(\partial_{jx}u_x)\frac{\delta}{\delta u_x}x_i(\frac{\delta}{\delta u_y}) h u_y - x_i(\frac{\delta}{\delta u_y}) h u_y (\partial_{jx}u_x)\frac{\delta}{\delta u_x}\} \\
    &=i\int dxdy \{-(\partial_{jx}u_x)\delta(x-y)x_ih(\frac{\delta}{\delta u_y}) + x_ih u_y (\partial_{jx}\delta(x-y))\frac{\delta}{\delta u_x}\} =i\int dx\{x_i\frac{\delta}{\delta u_x}h\partial_{jx}u_x+(\partial_{jx}x_ihu_x)\frac{\delta}{\delta u_x}\}=-i\delta_{ij}\hat{H}_2,\\
    &\int dx [x_i\hat{\mathcal{H}_4}, \hat{P}_j]=-i\int dxdy \{x_i\frac{\delta}{\delta u_y} h\frac{\delta}{\delta u^\dagger_y}(\partial_{jx}u^\dagger_x\frac{\delta}{\delta u_x^\dagger }+\partial_{jx}u_x\frac{\delta}{\delta u_x})\}
    =-i\int\{\partial_{jx}(x_i\frac{\delta}{\delta u_x}h)\frac{\delta}{\delta u^\dagger_x}+\frac{\delta}{\delta u_x}(\partial_{jx}x_ih\frac{\delta}{\delta u_x^\dagger})\}\\
    &=-i\int dx\{-x_i\frac{\delta}{\delta u_x}h\partial_{jx}\frac{\delta}{\delta u^\dagger_x}+\delta_{ij}\frac{\delta}{\delta u_x}h\frac{\delta}{\delta u_x^\dagger}+x_i\frac{\delta}{\delta u_x}h\partial_{jx}\frac{\delta}{\delta u^\dagger_x}\}=-i\delta_{ij}\hat{H}_4.
\end{align*}
Again, the proof of $\int dx [x_i\hat{\mathcal{H}_3}, \hat{P}_j]=-i\delta_{ij}\hat{H}_3$ is not shown above, since it is similar to the proof of $\int dx [x_i\hat{\mathcal{H}_2}, \hat{P}_j]=-i\delta_{ij}\hat{H}_2$. Combining all these identities, we obtain
\begin{equation*}
    [\hat{K}_i, \hat{P}_j] = -i\delta_{ij}( \frac{\lambda}{4}\hat{H}_1+\frac{1}{2}\hat{H}_2+\frac{1}{2}\hat{H}_3+\frac{1}{\lambda}\hat{H}_4) = -i\delta_{ij}\hat{H}.
\end{equation*}
The proofs of the rest of commutators for the Poincar\'{e} algebra in \eqref{PoincareAlgebra} are not shown here since they are very similar to the proofs shown in this Appendix.

\end{document}